\documentclass{aa}
\usepackage{graphicx}
\usepackage{float}
\usepackage{longtable}
\begin{document} 

   \title{Dust attenuation in z $\sim$ 1 galaxies from Herschel and 3D-HST H$\alpha$ measurements}
   \author{A. Puglisi
          \inst{1}
          \and
          G. Rodighiero\inst{1}
		  \and
		  A. Franceschini\inst{1}
		  \and
		  M. Talia\inst{2,3}
		  \and
		  A. Cimatti\inst{2}
		  \and
		  I. Baronchelli\inst{1}
		  \and
		  E. Daddi\inst{8}
		  \and	 
		  A. Renzini \inst{4}
		  \and 
		  K. Schawinski\inst{5}          
		  \and 
		  C. Mancini\inst{1,4}
		  \and 
		  J. Silverman\inst{10}
		  \and
		  C. Gruppioni\inst{3}
		  \and 		  
		  D. Lutz \inst{6}
	      \and
	      S. Berta \inst{7} 	
	      \and 
	      S. J. Oliver \inst{9}  
		  }

   \institute{Dipartimento di Fisica e Astronomia, Universit\`a di Padova, 
              vicolo dell'Osservatorio 2, I-35122, Padova, Italy\\
              \email{annagrazia.puglisi@studenti.unipd.it}
         \and
             Dipartimento di Fisica e Astronomia, Universit\`a di Bologna, 
             Via Ranzani 1, I-40127, Bologna, Italy \\
         \and
         	 INAF-Osservatorio Astronomico di Bologna,  
         	 Via Ranzani 1, I-40127, Bologna, Italy \\
		\and	         
         	 INAF-Osservatorio Astronomico di Padova,  
         	 Vicolo dell’Osservatorio, 5, I-35122, Padova, Italy
         \and
         	 Institute for Astronomy, Departement of Physics, ETH Zurich,  
         	 Wolfgang-Pauli-Strasse 27, CH-8093 Zurich, Switzerland \\ 
         \and
         	 Max-Planck-Institut für extraterrestrische Physik,  
         	 Giessenbachstrasse, D-85748 Garching, Germany \\
         \and
         	Max-Planck-Institut für Extraterrestrische Physik (MPE),  
         	Postfach 1312, 85741, Garching, Germany
		 \and         	
         	Laboratoire AIM-Paris-Saclay, 
         	CEA/DSM-CNRS-Universit\'{e} Paris Diderot, Irfu/Service d’Astrophysique, CEA
            Saclay, Orme des Merisiers, F-91191 Gif sur Yvette, France
          \and 
            Astronomy Centre, Dept. of Physics \& Astronomy, 
            	University of Sussex, Brighton BN1 9QH, UK
          \and 
          	Kavli Institute for the Physics and Mathematics of the Universe, 
          	Todai Institute for Advanced Study, the University of Tokyo, Kashiwa, Japan 277-8583 (Kavli IPMU, WPI)  	}

   \date{3$^{th}$ May 2015}
\abstract
{We combined the spectroscopic information from the 3D-HST survey with \textit{Herschel} data to characterize the H$\alpha$ dust attenuation properties of a sample of 79 \textbf{main sequence star-forming} galaxies at \textbf{$z \sim 1$} in the GOODS-S field. The sample was selected in the far-IR, at $\lambda$=100 and/or 160 $\mu$m, and only includes galaxies with a secure H$\alpha$ detection (S/N$>$3).
From the low resolution 3D-HST spectra we measured \textbf{the redshifts} and \textbf{the H$\alpha$ fluxes} for the whole sample (a factor of \textbf{1/1.2} was applied to the observed fluxes to remove the [NII] contamination).
The stellar masses (M$_{\star}$), infrared (L$_{IR}$) and UV luminosities (L$_{UV}$) were derived from the SEDs by fitting multi-band data from GALEX near-UV to SPIRE 500 $\mu$m.
We estimated the continuum extinction E$_{star}$(B-V) from both the IRX=L$_{IR}$/L$_{UV}$ ratio and the UV-slope, $\beta$, and found an excellent agreement between the two.
The nebular extinction was estimated from comparison of the observed SFR$_{H\alpha}$ and SFR$_{UV}$. 
We obtained \emph{f}=E$_{star}$(B-V)/E$_{neb}$(B-V)=0.93$\pm$0.06, i.e. higher than the canonical value of \textbf{\emph{f}=0.44} measured in the local Universe. 
Our derived dust correction produces good agreement between the H$\alpha$ and IR+UV SFRs for galaxies with SFR$\gtrsim$ 20 M$_{\odot}$/yr and M$_{\star} \gtrsim 5 \times 10^{10}$ M$_{\odot}$, while objects with lower SFR and M$_{\star}$ seem to require a smaller \emph{f}-factor (i.e. higher H$\alpha$ extinction correction).
Our results then imply that the nebular extinction for our sample is comparable to that in the optical-UV continuum and suggest that the \emph{f}-factor is a function of both M$_{\star}$ and SFR, in agreement with previous studies.}


\maketitle
\section{Introduction}
\label{introduction}
The rate at which a galaxy converts gas into stars, the Star Formation Rate (SFR)\textbf{,} is a fundamental quantity to characterize the evolutionary stage of the galaxy. The SFR can be measured in several ways, 
e.g. from the UV luminosity, the far-IR emission, the recombination lines or the radio-continuum \citep[]{Kenni98, Madau14}, although each of these SFR tracers suffers from some uncertainties. Concerning the UV and optical indicators, the main source of uncertainty \textbf{in the estimate of the SFR} is due to the dust extinction, that strongly absorbs the flux emitted by stars at UV and optical wavelengths and re-emits it in the far-IR. Hence, an accurate quantification of the impact of dust on the galaxy integrated emission is crucial to precisely evaluate the SFR.
\textbf{ The dust distribution inside a galaxy can be described by a two component model \citep[e.g.]{CharlotFall}, including a diffuse, optically thin component (the Inter-Stellar Medium, ISM) and an optically-thick one (the \emph{birth cloud}) related to the star forming regions.
The birth clouds have a finite lifetime ($\tau_{BC} \sim 10^7$ yr) and consist of an inner HII region ionized by young stars and bounded by an outer HI region.
This model assumes that the stars are embedded in their birth clouds for some time and then disrupt them or migrate away into the ambient ISM of the galaxy. In this model the emission lines are produced only in the HII regions of the birth clouds as the lifetime of the birth clouds is in general greater than the lifetimes of the stars producing most of the ionizing photons ($\sim 3 \times 10^{6}$ yr).
The emission lines and the non ionizing continuum from young stars are attenuated in the same way by dust in the outer HI envelopes of the birth clouds and the ambient ISM but, since the birth clouds have a finite lifetime, the non ionizing continuum radiation from stars that live longer than the birth clouds is attenuated only by the ambient ISM \citep{CharlotFall}.}

This two component model was conceived to explain the higher extinction observed on the nebular lines with respect to the UV/optical stellar continuum in the local Universe \citep[e.g.][]{Fanelli88, Hesse99, Mayya04, CidFernandes05, Calzetti94, Calzetti00}.

The relation between the color excess of the nebular regions and that of the stellar continuum derived by \cite{Calzetti94, Calzetti00}, i.e. \emph{f}=E$_{star}$(B-V)/E$_{neb}$(B-V)=0.44, has proven to be successful for local star-forming galaxies, while it is still unclear if it holds true in the high redshift Universe. 

The most secure method to quantify the amount of dust extinction in the HII regions is by directly measuring the Balmer decrement (i.e. the H$\alpha$/H$\beta$ line ratio). 

Nevertheless, such measurements are very challenging at $z \gtrsim $ 0.5 since the H$\alpha$ line is shifted to the less-accessible near-infrared window so indirect methods are often needed to infer the attenuation related to the nebular lines.

Current studies on dust properties in high redshift galaxies lead to contrasting results.
In fact, while some authors claim that the extinction related to the emission lines and to the stellar continuum are comparable in $z \sim 2$ galaxies \citep[e.g.][using UV-selection]{Erb06, Reddy10}, other authors confirm the validity of the Calzetti et al. local relation \citep[e.g][for samples of mainly optical and/or near-IR selected sources]{Schreiber, Whitaker, Yoshi}. Some recent works also suggested that the factor 
\emph{f}=E$_{star}$(B-V)/E$_{neb}$(B-V) 

required to reconcile the various SFR measurements is higher than those computed in the local Universe \citep[e.g.][on sBzk selections]{Kashino13, Pannella14}.
These contrasting results found in the literature are not surprising, given the different and indirect methods used and/or the small and sometimes biased samples of most of these studies. 
\textbf{Direct measurements of the Balmer Decrement would be required on statistical samples of sources at high redshift to clarify dust properties of distant star forming galaxies.}

In the present paper we used near-infrared spectroscopic data from the 3D-HST survey \citep{3dhst} and multi-wavelength photometry to derive the differential attenuation on a sample of 79 star forming galaxies at $z $ between 0.7 and 1.5, selected in the far-IR within the GOODS-South field. 
The presence of the far-IR photometry for the whole sample is very important, as these data allow us to robustly constrain the integrated dust emission. \textbf{This kind of approach has been often applied to galaxies in the local Universe \citep[see for example][]{DominguezSanchez14}, but only very occasionally at higher redshifts}. 

The paper is organized as follows. In Section \ref{Sample_section} the properties of the sample and the data-set are described. Section \ref{spectral_analysis} presents the spectral analysis. Section \ref{section4} illustrates the computation of the physical quantities for the galaxies in the sample. Section \ref{dustExt} describes the measurement of dust extinction on the H$\alpha$ emission. Section \ref{caveats} presents a critical discussion about the results whit carefully attention to the assumption behind our analysis and Section \ref{Conclusions} summarizes the results.

We adopt throughout this work a standard cosmology ($H_{0}=70 km/s/Mpc,\: \Omega_{m}=0.3, \: \Omega_{\Lambda}=0.7$) and assume a \cite{Salpeter} IMF.

\section{Sample and data-set}
\label{Sample_section}

Our sample is selected in the far-IR using the \textit{Herschel} observations from the PACS Evolutionary Probe survey \citep[PEP,][]{Lutz} in the GOODS-South field, that has the deepest sampling with PACS-\textit{Herschel} photometry.
The far-IR data are a key ingredient for the purpose of this work, as they allow us to strongly constrain the thermal emission by dust and therefore to infer  robust estimates of the continuum dust attenuation for all the galaxies in the sample.
The \textit{Herschel} selected sample has been matched with objects at $0.7 \textless z \textless 1.5$ from the 3D-HST survey catalog \citep{3dhst, Skelton} for which the H$\alpha$ line is observable with the G141 grism.
For the cross-correlation procedure we took advantage of the GOODS-Multiwavelength Southern Infrared Catalog \citep[\emph{GOODS-MUSIC},][]{Grazian}, that collects the available photometry \textbf{from $\sim$ 0.3 to 8 $\mu$m} for the objects detected in the GOODS-S field.
In the following we briefly describe the catalogs used and the sample selection procedure.

\subsection{The \textit{Herschel}/PEP survey}

The PEP survey is a deep extra-galactic survey based on the observations of the PACS instrument at 70, 100 and 160 $\mu$m. The GOODS-South field is the deepest field analyzed by the PEP survey and it is the only one observed also at 70 $\mu$m. The 3$\sigma$ limit in the GOODS-S field is 1.0 $mJy$, 1.2 $mJy$ and 2.4 $mJy$ at 70, 100 and 160 $\mu$m respectively.

\textbf{PACS photometry has been performed with a PSF fitting tool, adopting the positions of \textit{Spitzer} MIPS 24 $\mu$m detected sources as priors. This approach has been applied to maximize the depth of the extracted catalogs, to improve the deblending at longer wavelengths, and to optimize the band-merging of the \textit{Herschel} photometry to the available ancillary data in the UV-to-near IR. As described in \cite{Berta11, Berta13}, the 24 $\mu$m have been used as a bridge to match \textit{Herschel} to \textit{Spitzer}/IRAC (3.6 to 8.0 $\mu$m) and then to the optical bands. PACS priors source extraction followed the method described by \cite{Magnelli09}. The completeness of the catalogs have been estimated through extensive Monte-Carlo simulations and it turns to be on the order of 80\%, with flux limits of 1.39, 1.22 and 3.63 mJy at  70, 100 and 160 $\mu$m, respectively (see PEP full public data release \footnote{http://www.mpe.mpg.de/ir/Research/PEP/DR1}). We defer to \cite{Lutz} for further information on the PACS source extraction performances.}

\subsection{Multi-wavelength photometry}
\subsubsection{The GOODS-MUSIC catalog}

The GOODS-MUSIC catalog reports photometric data \textbf{for the sources detected in \emph{z} and \emph{Ks} bands in the GOODS-South area and it is entirely based on public data}. 
The catalog was produced with an accurate PSF matching for space and ground-based images of different resolutions and depths \textbf{and it includes 14847 objects}. The photometric catalog was cross-correlated with a master catalog\footnote{www.eso.org/science/goods/spectroscopy/CDFS\_Mastercat} released by the ESO-GOODS's team, that summarizes all the information \textbf{about the spectroscopic redshifts collected from several surveys in the GOODS area}. For the sources lacking a spectroscopic measurement, \textbf{\cite{Grazian} computed a photometric redshift using a standard $\chi^2$ technique with a set of synthetic templates drawn from the PEGASE2.0 synthesis model \citep{Fioc97}.}
The MUSIC catalog was extended by \cite{Santini09} with the inclusion of the mid-infrared fluxes, obtained from MIPS observations at 24 ${\mu}$m. 
This catalog was also used for the computation of the Spectral Energy Distributions (SED, see Section \ref{SEDfittingSection} for more details about the SED-fitting).

\subsubsection{GALEX, IRS-\textit{Spitzer} and SPIRE observations}
\label{GALEX,spitzer,SPIRE}

In order to increase the spectral coverage of the final sample, we added to the MUSIC photometry also the GALEX near-UV observations at $\lambda$ = 2310 {\AA} acquired from the online catalog Mikulski Archive for Space Telescopes (MAST), the IRS observations at 16 ${\mu}$m \citep{Teplitz} and the SPIRE data at 250, 350 and 500 ${\mu}$m (\citealt{Oliver12, Roseboom10, Levenson10}, see also \citealt{Viero13} and \citealt{Gruppioni13}).

Both the GALEX and the SPIRE observations are extremely important for the scope of this work. The GALEX data \textbf{allow} us to better constrain the SED at UV wavelengths, enabling \textbf{a more} accurate estimate of the UV spectral slope $\beta$ (see Section \ref{continuum attenuation} and Appendix \ref{AppendixA} for details)\textbf{. On the other hand,} the SPIRE data improve the spectral coverage in the far-IR, thus \textbf{allowing us} to derive a robust estimate of the bolometric infrared luminosity (see Section \ref{SEDfittingSection}).

\textbf{The SPIRE catalog is also selected with positional priors at 24 $\mu$m. 
SPIRE fluxes were obtained with the technique described in \cite{Roseboom10}. Images at all Herschel wavelengths have undergone extractions using the same MIPS prior catalog positions.}

\textbf{For the GALEX data we used publicly available photometric catalog from MAST \footnote{http://archive.stsci.edu/index.html}, 
and rely in this case on a simple positional association.}

\textbf{The ancillary photometry was included by cross-correlating the above mentioned catalog using a positional association with a matching radius of 1 $arcsec$.}
About 28 \% of the galaxies in the final sample have a near-UV observation, the 82 \% of these sources have a 16 $\mu$m counterpart, while 87 \%, 62 \% and 23 \% of the galaxies have a SPIRE 250, 350 and 500 $\mu$m detection, respectively. \textbf{This fast decreasing fraction with $\lambda$ is due to the degrading of the PSF in the SPIRE diffraction-limited imager}.

\subsection{The 3D-HST survey}
\label{3dhstsection}

The 3D-HST near-infrared spectroscopic survey \citep{3dhst, Skelton} covers roughly 75\% of the area imaged by the CANDELS ultra-deep survey fields\citep{CandelsA, CandelsB}. In this work we used the observations in the GOODS-S field from the preliminary data release, that covers an area of about 100 square $arcminutes$.
The WFC3/G141 grism is the primary spectral element used in the survey. The G141 grism covers a wavelength range from 1.1 to 1.65 ${\mu}$m, thus can detect the H$\alpha$ emission line in a redshift interval between 0.7 and 1.5. The mean resolving power of this grism is about $R \sim 130$, insufficient to deblend the H$\alpha$ and [NII] emissions. \textbf{Our measured H$\alpha$ flux is then affected by contamination from [NII], that we took into account and removed as described in Section \ref{SFRsection}.}

Because no slit is used, and the length of the dispersed spectra is larger than the average separation of galaxies down to the detection limit of the survey \textbf{(F$_{obs}(\lambda)$ $\sim$ $2.3 \times 10^{-17} erg/s/cm^2$ at 5$\sigma$)}, there is a significant chance that the spectra of nearby objects could be overlapped. This ``contamination'' by the neighbors must be carefully accounted for in the analysis of the grism spectra. \textbf{To evaluate the contribution to the flux from nearby sources, we considered the quantitative model developed by \cite{3dhst}. 
For a complete description of the instrumental set-up, the data-reduction pipeline and the contamination model we defer to \cite{3dhst}.
Further details about the contamination issue are discussed in Sections \ref{cleaning} and \ref{spectral_analysis}.}

\subsection{Cross correlation and cleaning of the sample}
\label{cleaning}

The use of the MUSIC catalog is fundamental for the association between the \textit{Herschel} and 3D-HST observations\textbf{, as} the \textit{Herschel}'s beam is not directly matchable to the high resolution imaging of HST: the spatial resolution of PACS at short and long wavelengths is $\sim 5$ $arcsec$ and $\sim 11$ $arcsec$ respectively, while the WFC3 camera has a spatial resolution of 0.13 $arcsec$. The inclusion of the MIPS data at 24 ${\mu}$m was very useful for the association with the \textit{Herschel} data, due to the \textbf{\textit{Spitzer} - MIPS} intermediate resolution between optical instruments and the \textit{Herschel} Space Telescope. 

The PACS-MUSIC catalog includes 591 sources, selected at 100 and/or 160 $\mu$m above the 3$\sigma$ flux limits of 1.1 $m$Jy and 2 $m$Jy, respectively. About 40\% of these objects have ground-based spectroscopic redshift in the MUSIC catalog.
We then cross-correlated the PACS-MUSIC and 3D-HST catalogs, using a matching radius of 2 $arcsec$, and found that 378 PACS-MUSIC objects have a counterpart in the 3D-HST catalog.
\textbf{The choice of the adopted near-search radius, 2 $arcsec$, to cross-match the PACS-MUSIC sample and the 3D-HST is the best compromise between the 24 $\mu$m beam size (5.6 $arcsec$) and the IRAC ones (from 1.7 to 2 $arcsec$ going from the 3.6 $\mu$m to the 8.0 $\mu$m channel). 
We have however verified that using a slightly smaller (1 $arcsec$) or larger (3 $arcsec$) search radius does not affect the statistics of our final sample.}
\textbf{Of these 378 sources}, we only considered the 144 galaxies with a MUSIC redshift between 0.7 and 1.5, for which the H$\alpha$ emission line is expected to fall in the WFC3-G141 grism spectral range.
The spectra of these sources have been  visually inspected to discard faint or saturated objects or sources without \textbf{emission lines}. We excluded also sources that are strongly contaminated by nearby objects, \textbf{carefully evaluating by eye the 2D and 1D spectra of each galaxy, also considering the quantitative contamination model of \cite{3dhst}. Galaxies whose spectra are strongly contaminated were discarded from the final sample.} To avoid mis-identification of the lines, we also discarded objects with $\textbar z_{MUSIC} - z_{3D-HST} \textbar \geq 0.2$.

\subsubsection{\textbf{The X-ray luminous AGN population}}

We used the Chandra 4Ms catalog from \cite{Xue11} to identify and discard the AGN from our sample. Following \cite{Xue11}, we classified as AGN only those objects with an absorption-corrected rest frame 0.5-9 keV luminosity L$_{X-ray}$ $\geqslant$ $3 \times 10^{42}$ $erg/s$: we found that 8 objects match this criterion. We cannot confirm the presence of the AGN \textbf{relying} on the 3D-HST data alone, \textbf{as} they do not have neither the required resolution to deblend the [NII] and H$\alpha$ lines, nor the adequate spectral coverage to detect other emission lines entering the BPT diagram \citep{Baldwin81}, e.g. [OIII]$_{5007}$ and H$\beta$.

\subsection{The final sample}
\label{final_sample}

To summarize, since we wanted to study galaxies with detections in the far-IR and in the H$\alpha$ emission line, we considered only the 144 galaxies in the range $0.7 \leq z \leq 1.5$ from the full sample of PACS objects with 3D-HST counterparts. The spectra of these sources were then visually inspected to discard the faint, noisy, saturated or strongly contaminated ones, removing AGN as explained in the previous section.

\begin{figure}[h!!!]
\centering
\includegraphics[scale=0.5]{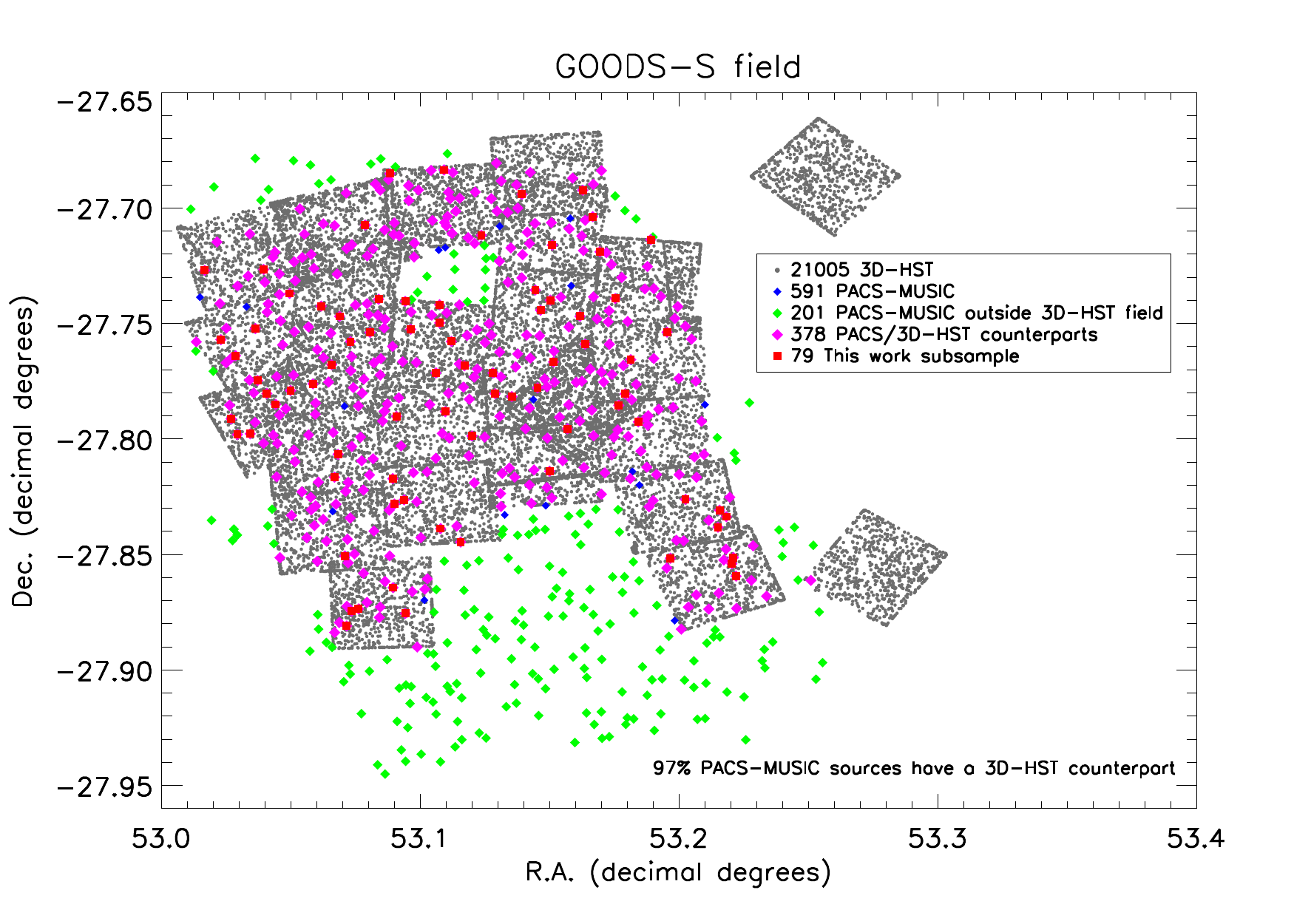}
\caption{Overlay showing the spatial distribution of the PACS/MUSIC/3D-HST source sample.}
\label{overlay}
\end{figure}

The final sample includes 79 sources, i.e. $\sim$55\% of the 144 PACS-selected objects at $0.7 \leq z \leq 1.5$ with a counterpart in the 3D-HST catalog. The sample galaxies have observed H$\alpha$ luminosities L$_{H\alpha,obs}$ $\in$ $[1.4 \times 10^{41} - 5.6 \times 10^{42}]$ $erg/s$ and bolometric infrared luminosity L$_{IR}$ $\in$ $[1.2 \times 10^{10} - 1.3 \times 10^{12}]$ L$_{\odot}$.
\textbf{An overlay showing the spatial distribution of our source sample is reported in Figure \ref{overlay}}.

\begin{figure}[h!!!]
\centering
\includegraphics[scale=0.5]{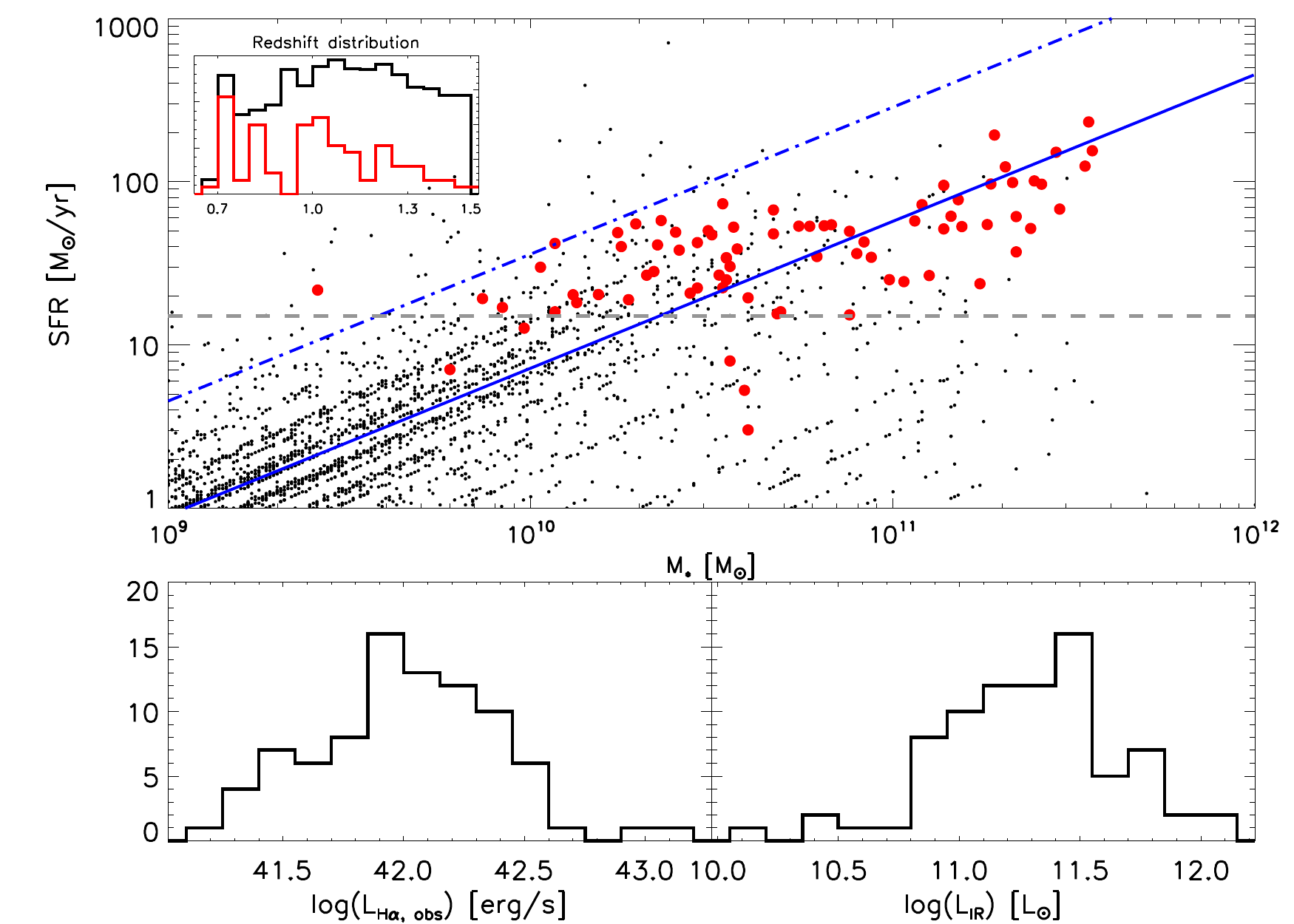}
\caption{Location of our sample (red filled circles) in the SFR-M$_{\star}$ space, compared with the distribution of the all 3D-HST galaxies (black \textbf{dots}) in the same redshift range.
The blue solid line represents the Main Sequence at $z \sim 1$ \citep{Elbaz07} and the dot-dashed blue line is the 4$\times$MS. The grey dashed line represents the SFR corresponding to the 3$\sigma$ flux limit at 160 $\mu$m ($f_{\lambda}$=2.4 $mJy$) for a typical main-sequence galaxy, derived using the median SEDs of \cite{Magdis12}.
The inset shows the redshift distribution for this work sample (red \textbf{curve}) compared to the complete 3D-HST sample in the same redshift range (black \textbf{curve}).
The lower panel on the left is the distribution of the observed H$\alpha$ luminosities for our sample (not corrected for dust attenuation), the lower panel on the right shows the distribution of L$_{IR}$.}
\label{parent_sample}
\end{figure}

Due to the adopted selection criteria, our sources are not fully representative of the whole Main Sequence (MS) population at $z \sim 1$ \citep{Elbaz07, Noeske07}. This effect is visible in Figure \ref{parent_sample}, showing the distribution of our sample in the M$_{\star}$-SFR plane with respect to the 3D-HST galaxies at $z \in [0.7 - 1.5]$. The stellar masses and SFR for the overall 3D-HST population are derived from SED fitting \citep{Skelton}. The SFR for the galaxies analyzed in this work are instead estimated from the IR+UV luminosities and the stellar masses M$_{\star}$ are computed using the MAGPHYS software \citep{Magphys}, as detailed in Sections \ref{SEDfittingSection} and \ref{SFRsection}.
Fig. \ref{parent_sample} offers a qualitative comparison of the two selections, and has the only purpose of showing the effects of the used selection criteria on the global properties of our sample.
Our objects occupy the upper part of the MS of ``normal'' star-forming galaxies at $z \sim 1$, due to the adopted far-IR selection \citep[correspondng to a selection in SFR, cfr.][]{Rodighiero14}. Only one galaxy lies 4$\times$above the MS, i.e. the starburst region \citep{Rodighiero11}. However, the presence of this outlier does not influence our results, hence we did not remove it from the sample. The SFR and mass ranges spanned by our sample are $3 \leq$ SFR$_{IR+UV} \leq 232$ M$_{\odot}$/yr, $2.6 \times 10^9 \leq$ M$_{\star} \leq 3.5 \times 10^{11}$M$_{\odot}$, respectively.

\begin{figure}[h!!!]
\centering
\includegraphics[scale=0.5]{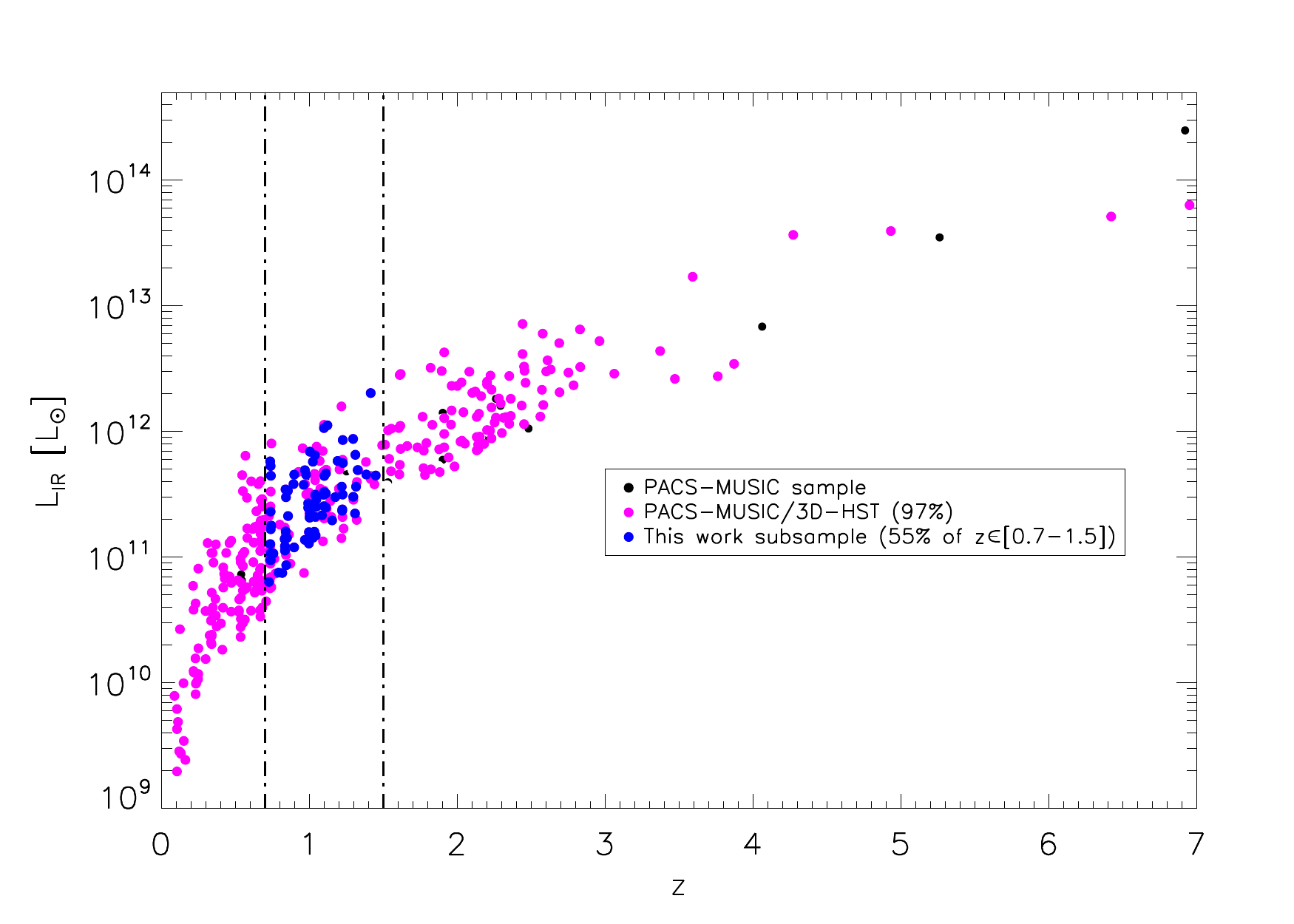}
\caption{Total infrared luminosity L$_{IR}$ as a function of redshift for the PEP sample. The magenta filled circles are the 378 PEP sources with a counterpart in the 3D-HST observations. The blue filled circles highlight the 79 sources of our final sample, located in the redshift range $z \in [0.7 - 1.5]$ }
\label{Lirz}
\end{figure}

Figure \ref{Lirz} shows the dependence of the total infrared luminosity as a function of the redshift for the PACS-MUSIC sources that are located inside the area observed also by the 3D-HST survey:  97 \% of these objects have a 3D-HST counterpart. In Figure \ref{Lirz} the 79 sources of \textbf{our final} sample, namely the PACS-MUSIC/3D-HST sources with an H$\alpha$ emission and a ``clean'' spectrum, are also highlighted.

\section{Spectral analysis}
\label{spectral_analysis}

\begin{figure}[h!!!]
\centering
\resizebox{\hsize}{!}{\includegraphics[scale=0.9]{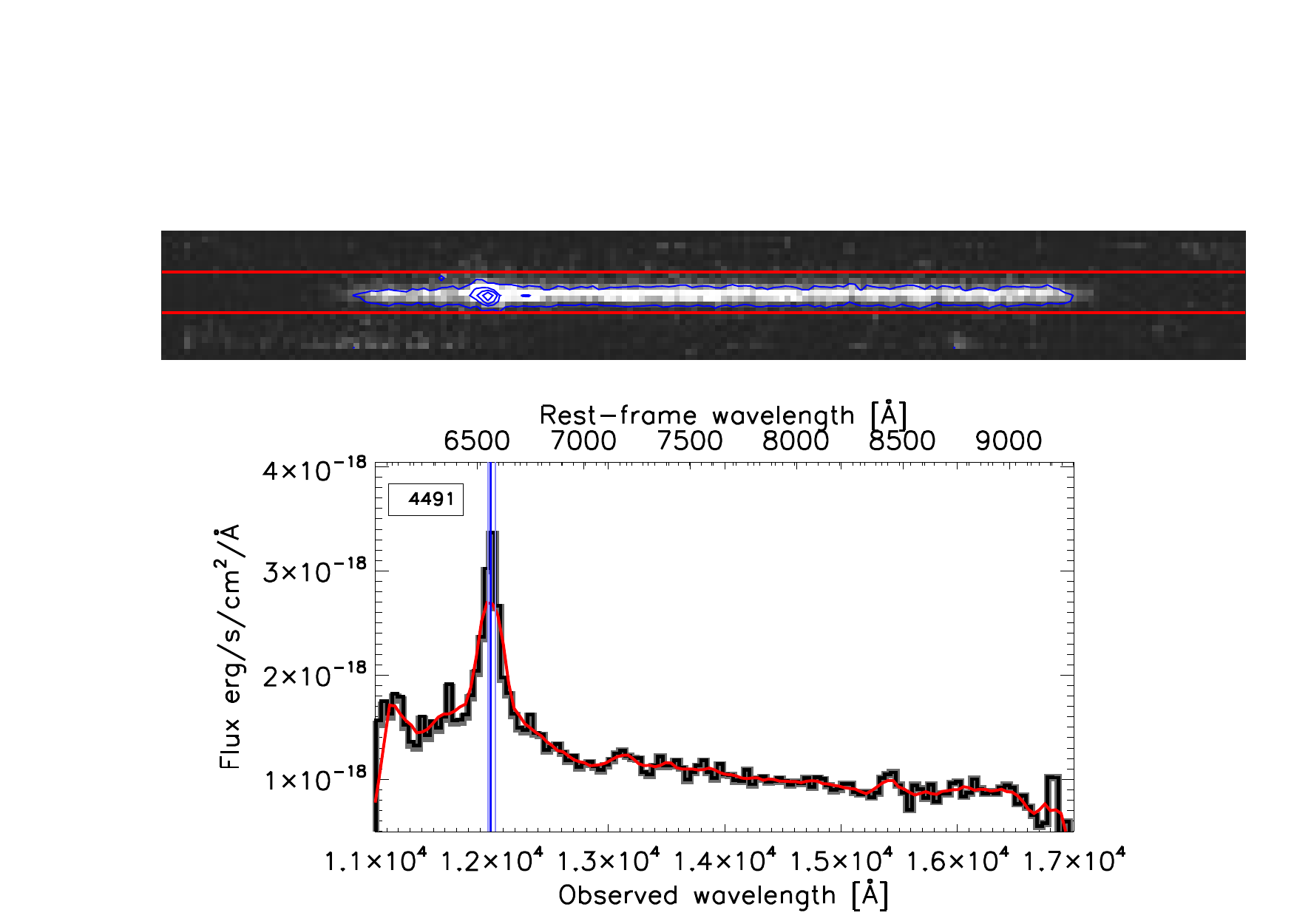}}
\caption{\emph{Upper panel:} image of the 3D-HST bi-dimensional spectrum for source 4491 (ID-MUSIC), as displayed by the IDL routine. The red horizontal lines define the ``virtual slit''. 
\emph{Lower panel:} 1D spectrum of source 4491, obtained collapsing the 2D spectrum along the columns inside the virtual slit. The flux is in $erg/sec/cm^2/${\AA}.
The black thicker histogram is the observed spectrum, the red curve is a spline interpolation to the observed spectrum. \textbf{The grey thin histogram, almost overlapping the black one, represents the "decontaminated spectrum", i.e. the contamination model by \cite{3dhst} subtracted to the observed spectrum. The overlap between the observed and the decontaminated spectrum highlight the fact that the contamination flux in the region defined by the virtual slit is negligible.}
The vertical blue line highlights the position of the H$\alpha$ emission while the pale blue lines correspond to the position of the [NII] doublet.}
\label{spectrum IDL}
\end{figure}

We extracted the 1D spectra by collapsing the 2D spectra inside specific apertures, adapting an IDL procedure which also applies the calibration files by \cite{3dhst}. Figure \ref{spectrum IDL} shows the grism spectrum and the extracted 1D spectrum for one of the sources in our sample.
The upper panel displays the cutout of the grism exposure: the 2D spectrum of the object is located in the central part of the frame.

The extraction of the 1D spectrum (lower panel of Figure \ref{spectrum IDL}) is done inside a frame region called ``virtual slit'' (inside the two red horizontal lines in the upper panel of Figure \ref{spectrum IDL}). The width and the position of the virtual slit are adjusted ad hoc on each object to maximize the S/N ratio and to minimize the contribution to the flux from other spectra.
The flux contribution from other sources is negligible in the extracted spectra, as we defined the position of the virtual slits to minimize this effect. 
\textbf{Furthermore, for the majority of galaxies in our final sample, this confusion usually affects the flux at all wavelengths, so the contamination, if present, is mostly removed by the flux correction to the spectrum as discussed in Sect. \ref{ApCorrSection}.
For 9 sources out of 79, we found that the flux is strongly contaminated at the edge of the spectrum but the spectral range around the H$\alpha$ and the closeby continuum are not affected by this contamination. We accounted for this problem in these 9 sources by carefully selecting the portion of the continuum near the H$\alpha$ emission when rescaling the spectrum to the observed broad-band photometry (i.e. to derive the aperture correction factor) and excluding the part of the continuum affected by the contamination.}

\subsection{H$\alpha$ fluxes and redshift measurements}
\label{z_flux_measure_section}

\begin{figure}[h!]
\centering
\includegraphics[scale=0.5]{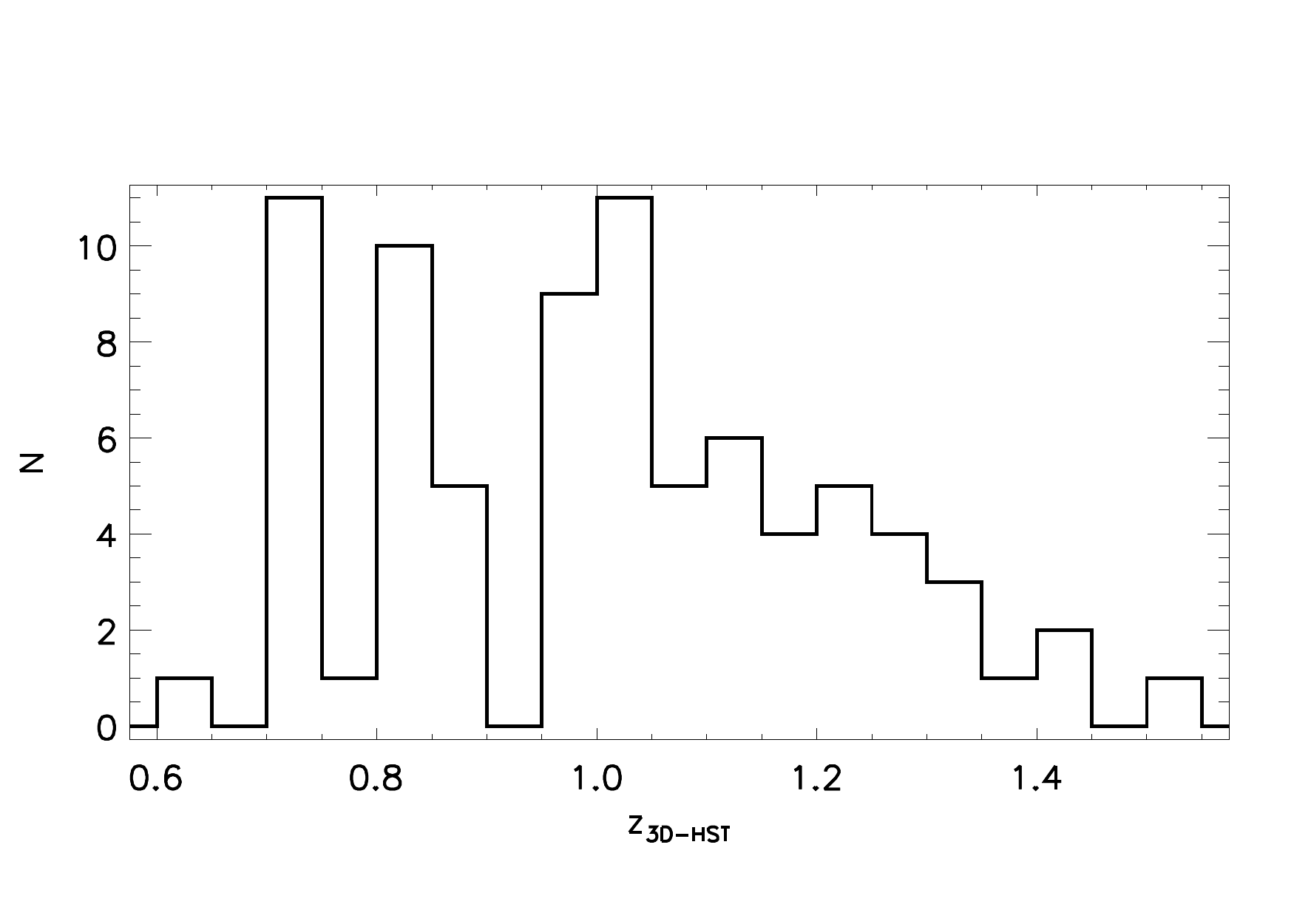}
\caption{Redshift distribution for the 3D-HST sources in the GOODS-South field. 
The sources are distributed in a redshift interval $z$ $\in$ $[0.65-1.53]$, according to the features of the WFC3/G141 grism.}
\label{zdis}
\end{figure}

\begin{figure}[h!]
\centering
\includegraphics[scale=0.5]{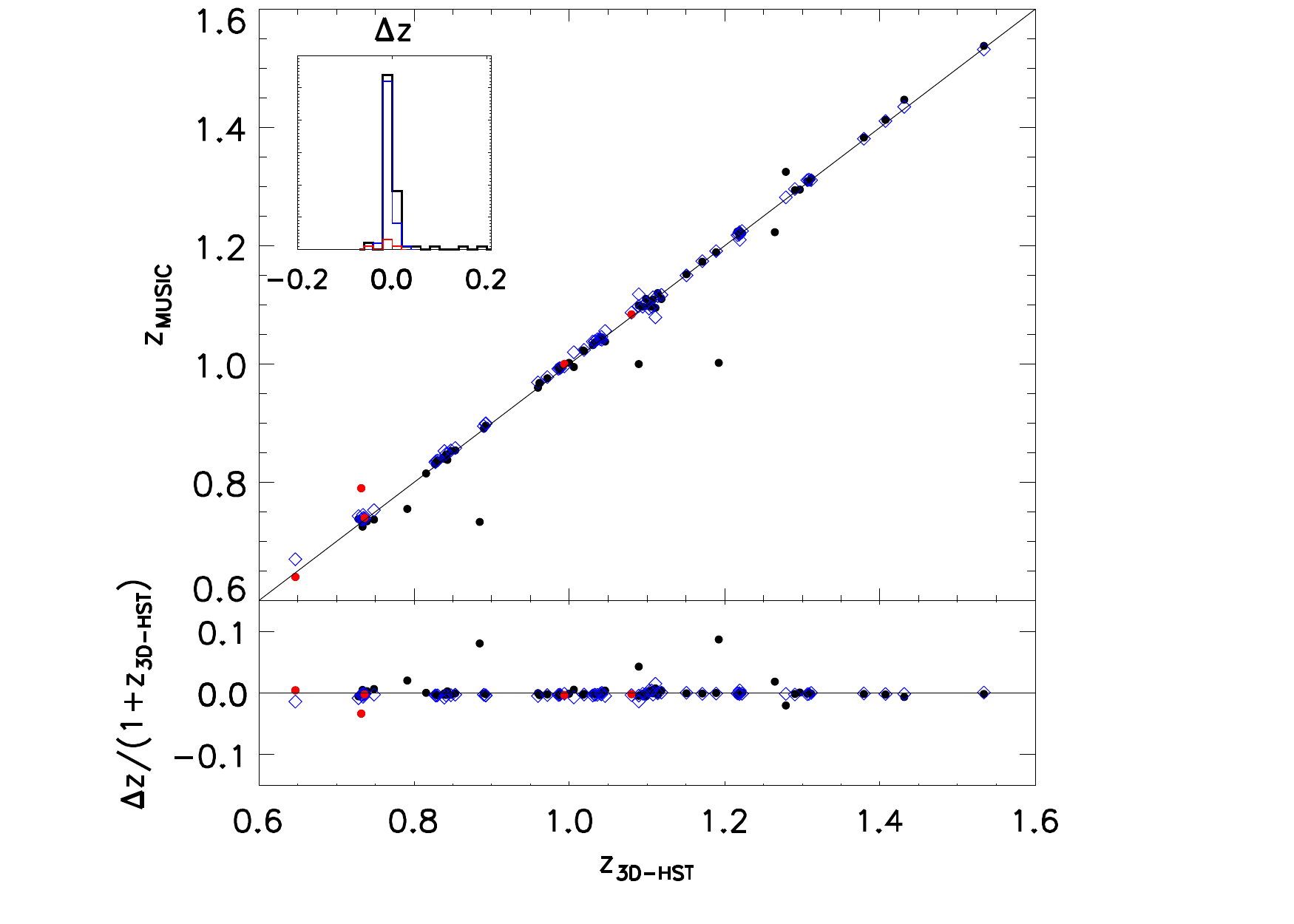}
\caption{\emph{Upper panel:} relation between 3D-HST spectroscopic redshifts (\emph{x}-axis) and MUSIC redshifts. 
In the inset, the distribution of the absolute scatter $\Delta z$=($z_{3D-HST}$-$z_{MUSIC}$) is shown. The standard deviation for this distribution is $\sigma$=0.031. The red histogram highlights the absolute scatter for photometric MUSIC redshifts. The blue histogram represents the absolute scatter between our measurements and the redshifts from \cite{Morris15}.
\emph{Lower panel:} relative scatter ($z_{3D-HST}$-$z_{MUSIC}$)/(1+$z_{3D-HST}$). 
The data points in red are the photometric redshifts in the MUSIC catalog, the black dots are the sources with spectroscopic redshift also in the MUSIC catalog from ground based measurements. The blue open diamonds are the redshifts measured by Morris et al..}
\label{zcorr}
\end{figure}

We measured the redshift and the integrated H$\alpha$ flux from the extracted 1D spectra by fitting the emission lines with a Gaussian profile based on the IRAF tool \textit{splot}, even if the spectral line profile is dominated by the object shapes \citep[the so-called ``morphology'' broadening, ][]{Schmidt13} and the grism line shape may not be well described by the Gaussian profile. 
\textbf{For part of the sample sources we re-measured the fluxes by simply integrating the area under the line, without fitting any profile, and found that the flux measurements are in agreement with those obtained with the Gaussian fit.}
The errors on the redshift and on the H$\alpha$ integrated flux were estimated using Monte Carlo simulations with random Gaussian noise. 
The measurements of the observed H$\alpha$ luminosities and the redshifts are listed in Tab. \ref{data_sample} in Appendix \ref{appendix_spectra}.
\\The redshifts are distributed between $z$ $\sim$ 0.7 and 1.5 as shown in Figure \ref{zdis}, as a result of the combined transmission of the grism G141 + F140W filter.
Our measurements are in good agreement with the pre-existing redshift measurements.
Figure \ref{zcorr} shows the excellent correlation between the MUSIC spectroscopic (black dots) and photometric (red dots) redshifts and our redshifts measured from 3D-HST spectra. The median absolute scatter $\Delta z = \textbar z_{\footnotesize{3D-HST}} - z_{\footnotesize{MUSIC}} \textbar$ is $0.0040$, while the median relative scatter $\Delta z / (1 +  z_{\footnotesize{3D-HST}}) \simeq 0.0022$.
We also compare our redshifts with those from the recent catalog of \cite{Morris15}, measured also from 3D-HST spectra: also in this case a very good agreement is found ($\Delta z = \textbar z_{\footnotesize{3D-HST}} - z_{\footnotesize{Morris} \textbar} \simeq 0.0051$, $\Delta z / (1 +  z_{\footnotesize{3D-HST}}) \simeq 0.0027$).


\subsection{Flux correction by SED scaling}
\label{ApCorrSection}
The position and the width of the virtual slit used to extract the 1D spectrum 
were adjusted for each object in order to maximize the S/N ratio and minimize the contamination from nearby sources.
We applied corrections to the measured H$\alpha$ fluxes to account for flux losses outside the slit. 
For each 1D spectrum the correction factor was derived as the ratio between the average continuum flux in the 1D spectrum (F$_{spec}$) and on the galaxy SED (F$_{SED}$, see Sect. \ref{SEDfittingSection} for details about the SED fitting), measured in the same wavelength range.
\textbf{The average value of the correction factor is F$_{SED}$/F$_{spec}$ $\sim 1.41$, ranging from F$_{SED}$/F$_{spec} \sim$ 0.39 up to F$_{SED}$/F$_{spec} \sim$ 8.24} .

\begin{figure}[h!]
\centering
\resizebox{\hsize}{!}{\includegraphics[scale=1]{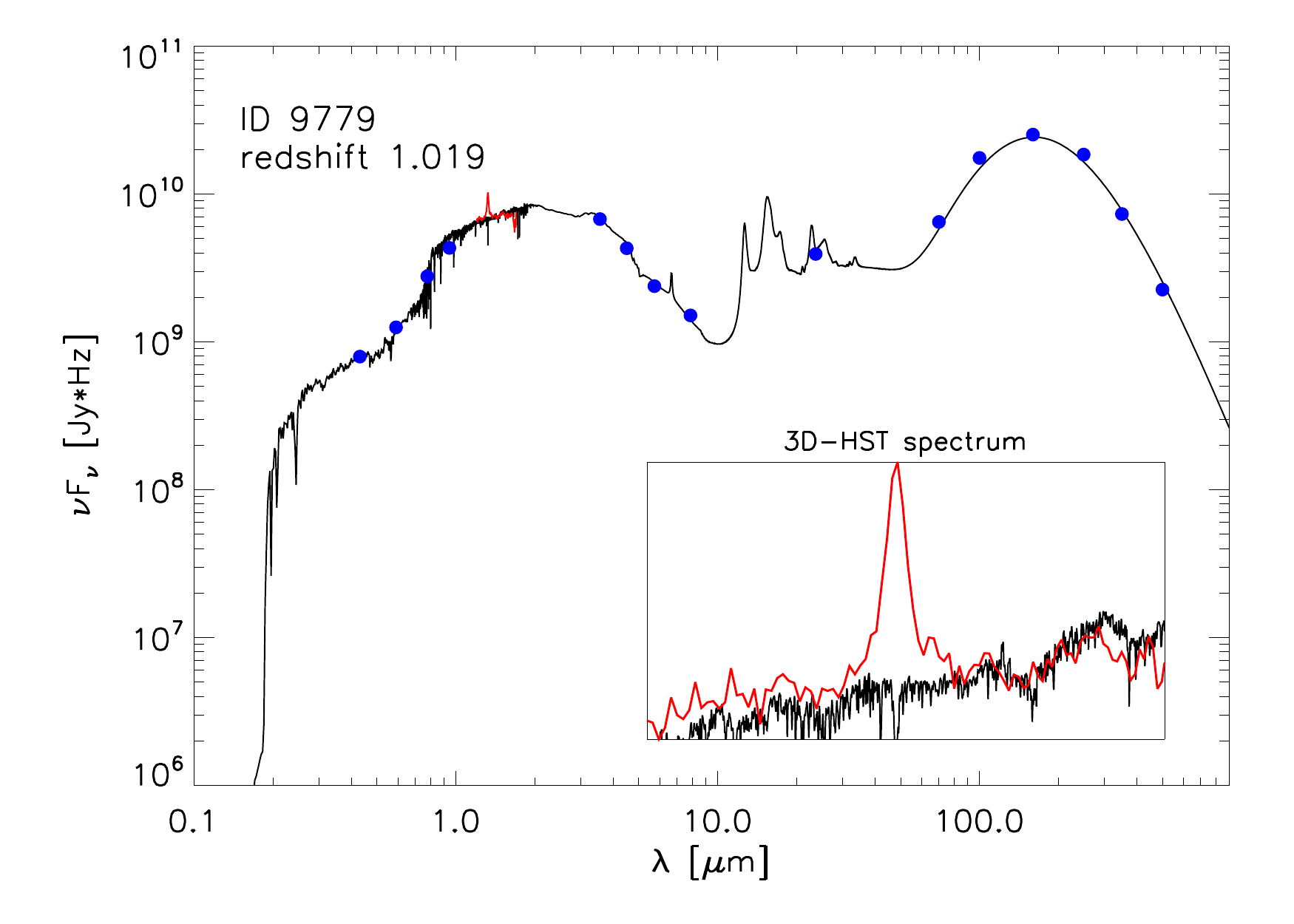}}
\caption{Spectral Energy Distribution from MAGPHYS (black curve) with the observed photometry (blue filled circles) and the 3D-HST spectrum (red line). The 3D-HST spectrum is corrected for the flux loss outside the slit, as described in the text.
The plot also reports the MUSIC ID of the source and its redshift, measured from the near-IR spectrum. The inset shows a zoom on the 3D-HST spectrum, close to the H$\alpha$ emission line.}
\label{apCorr_sed}
\end{figure}

Once this flux correction is applied, all the objects show an excellent agreement between the 1D spectrum and the near-IR part of the SED. An example is shown in Figure \ref{apCorr_sed}.

\section{Derivation of the main physical quantities of the sample galaxies}
\label{section4}

\subsection{SED-fitting}
\label{SEDfittingSection}

\begin{figure}[h!]
\centering
\resizebox{\hsize}{!}{\includegraphics[scale=1]{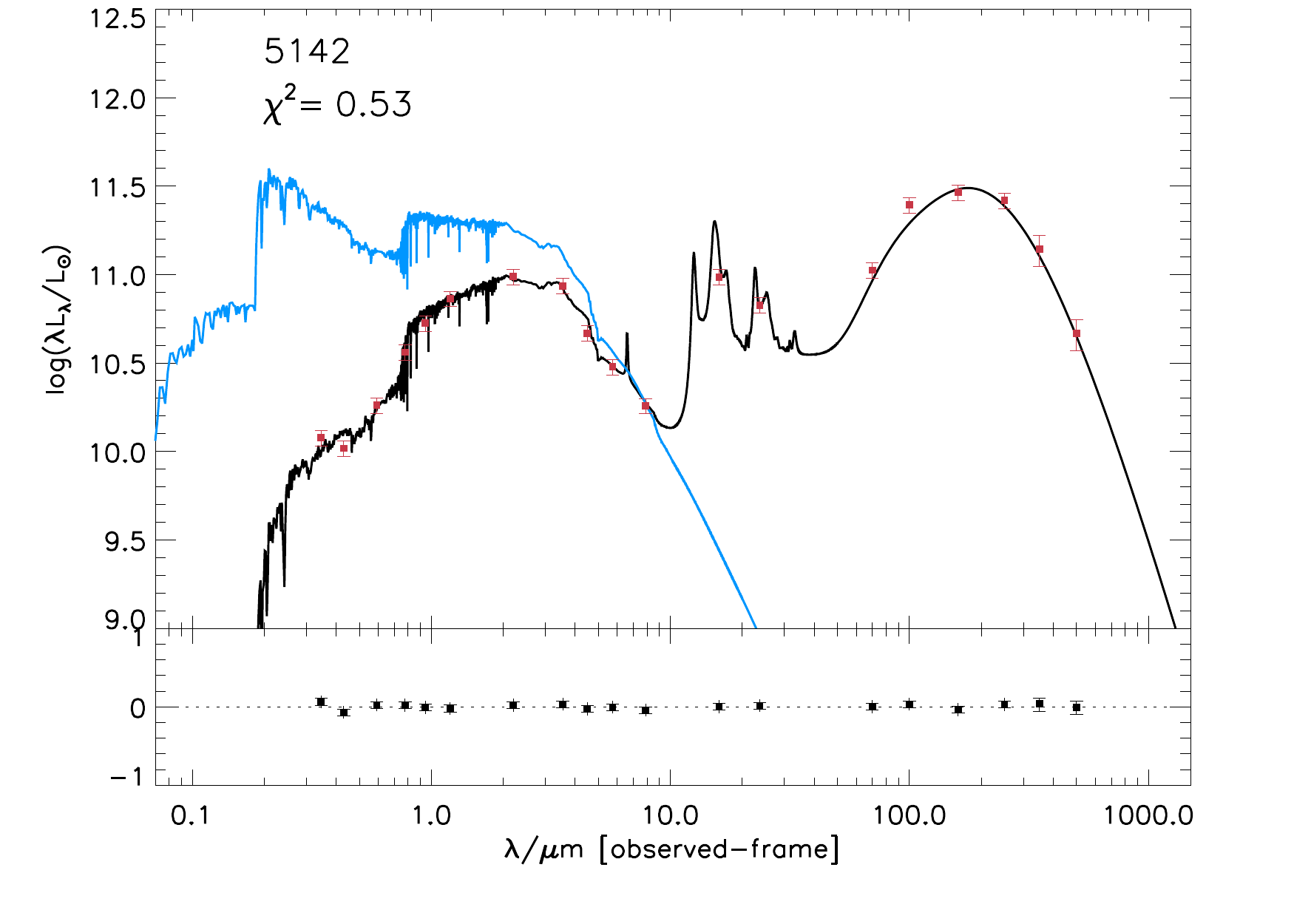}}
\caption{Example of a Spectral Energy Distribution in output from MAGPHYS. The black solid line is the best-fit model to the observed SED (data-points in red). The blue solid line shows the unattenuated stellar population spectrum. The bottom panel shows the residuals (L$_{\lambda}^{obs}$-L$_{\lambda}^{mod})$/L$_{\lambda}^{obs}$.}
\label{3636sed}
\end{figure}

We fitted the Spectral Energy Distributions with the MAGPHYS package \citep[Multi-wavelength Analysis of Galaxy Physical Properties, ][]{Magphys}.
This software has the main advantage of fitting the whole SED from the UV to the far-IR, relating the optical and IR libraries in a physically consistent way.
\textbf{To compute the SEDs, we ran MAGPHYS in the default mode, using the stellar-population synthesis models of \cite{BC03}.}
For the SED fitting we used 21 photometric bands, i.e. from the near-UV (GALEX) to the SPIRE 500 $\mu$m.
The filters used for the SED fitting are listed in Table \ref{filtersMagphys}. 
\textbf{The errors of the photometric data points were set to be 10\% of the measured flux and were then adjusted ad hoc.}
The best-fit SEDs were obtained by fixing the redshifts to the spectroscopic values that we measured from the 3D-HST spectra.
Figure \ref{3636sed} shows an example of a best-fit SED as output by MAGPHYS. 

In addition to the best-fit SED, MAGPHYS returns several physical parameters of the observed galaxy together with their marginalized likelihood distributions. 
We used in particular the stellar masses M$_{\star}$ and the bolometric infrared luminosity L$_{IR}$, considering the values at the 50$^{th}$ percentile of the likelihood distribution. \textbf{The errorbars at 68\% for these quantities were derived from the PDFs computed by MAGPHYS}.
Since MAGPHYS adopts a \cite{Chabrier} IMF, we converted the stellar masses to a Salpeter IMF by \textbf{multiplying} by a constant factor of 1.7 \citep[e.g][]{Cimatti08}.

\begin{center}
\begin{table}[h]
\caption{Filters used for the SED fitting and respective effective wavelengths $\lambda_{eff}$ ($\mu$m).}
\label{filtersMagphys}
\centering
\begin{tabular}{cc}
\hline \hline
Filter & $\lambda_{eff}$ [$\mu$m]  \\ \hline
GALEX\_NUV 	&	0.2310	\\
U	 		&	0.346	\\
ACSF435W 	&	0.4297 \\	
ACSF606W 	&	0.5907	\\
ACSF775W 	&	0.7774	\\
ACSF850LP	&	0.9445	\\
ISAACJ	 	&	1.2	\\
ISAACH	 	&	1.6	\\
ISAACK	 	&	2.2	\\
IRAC1	 	&	3.55	\\
IRAC2	 	&	4.493\\	
IRAC3	 	&	5.731	\\
IRAC4	 	&	7.872	\\
IRS16	 	&	16	\\
MIPS24	 	&	23.68\\	
PACS70		&	70	\\
PACS100	 	&	100\\
PACS160 		&	160	\\ 
SPIRE250		&	250 	\\
SPIRE350		&	350	\\
SPIRE500		&	500	\\\hline\hline
\end{tabular}
\end{table}
\end{center}

\subsection{Measuring the SFRs}
\label{SFRsection}

We derived for all the sources in the sample the SFR from several indicators, thanks to the availability of a wide multi-wavelength photometric coverage and the near-IR spectra. \textbf{In particular, we computed the SFRs from the H$\alpha$ luminosity, the UV luminosity and the IR luminosity, adopting the classical calibrations of \cite{Kenni98}}.

\textbf{The} H$\alpha$ luminosity is derived from 3D-HST near-IR spectra. The measured H$\alpha$ flux is aperture corrected as described in section \ref{ApCorrSection} while the [NII] contribution was removed by scaling down the H$\alpha$ flux by a factor 1.2, following \cite{Wuyts13}. 
The contamination by [NII] could produce some uncertainties in the estimate of L$_{H\alpha}$ since the ratio [NII]/H$\alpha$ may vary object by object as a function of the stellar mass and the redshift \citep{Zahid14}. 
\textbf{To validate our choice of using a constant factor, we then computed the [NII] correction factor as a function of M$_{\star}$, using the evolutionary parametrization of the ISM metallicity by \cite{Zahid13}. 
We found that the use of a more accurate [NII] correction does not change our results about the dust extinction corrections (i.e. Figures \ref{Betaerror} and \ref{compareSFR}). Results of our later analysis (like the \emph{f}-factor and the comparison between SFR indicators) are not affected by the choice of a mass-dependent [NII] correction or of a constant scaling factor.} 

In order to account for both the unattenuated and the obscured Star Formation, we measured the ``total'' SFR combining the L$_{IR}$ and the observed L$_{1600}$ \citep[the same procedure was adopted by][]{Nordon12}:
\begin{equation}
SFR_{IR+UV}=(1.4L_{1600} + 1.7L_{IR})\times 10^{-10}[L_{\odot}] \ .
\end{equation}
We estimated the luminosity at $\lambda_{rest-frame}=1600${\AA} from the best-fit SED, and used the IR luminosity L$_{IR}$ derived by MAGPHYS (cfr. Section \ref{SEDfittingSection}).
The SFR$_{IR+UV}$ is the best estimate of the SFR for our objects, considering the wide photometric coverage in the far-IR, and will be used as a benchmark to verify the validity of our dust extinction correction.
\\The errors for the various estimates of the SFR are computed via MC simulation by randomly varying L$_{H\alpha}$, L$_{1600}$ and L$_{IR}$ within the errorbars, assuming that these quantities are Gaussian distributed.

\section{Dust extinction corrections}
\label{dustExt}

In order to precisely evaluate the SFR and to reconcile the various SFR estimates, we need to accurately quantify the effect of dust on the integrated emission of our galaxies.
\textbf{According to \cite{RewCalzetti}, the action of dust on starlight for starburst galaxies in the local Universe can be parametrized as:}

\begin{equation}
F_{obs}(\lambda)=F_{int}(\lambda) \times 10^{-0.4A_{\lambda}}= 
 F_{int}(\lambda) \times 10^{-0.4E_{star}(B-V)k(\lambda)}
\label{flux}
\end{equation}

\textbf{where F$_{obs}(\lambda)$ and F$_{int}(\lambda)$ are the dust-obscured and the intrinsic stellar continuum flux densities, respectively. A$_{\lambda}$ and E$_{star}(B-V)$ are the dust attenuation and the color excess on the stellar continuum respectively, while $k(\lambda)$ parametrizes the \cite{Calzetti94} starburst reddening curve.}

\textbf{As already stated in the introduction,} \cite{Calzetti00} found that exists a differential attenuation between the stellar continuum and the nebular lines by measuring both the hydrogen line ratios and the continuum reddening in a sample of local star-forming galaxies. This differential attenuation is parametrized by a \emph{f}-factor of 0.44, as defined by :

\begin{align}
& E_{star}(B-V)=f \times E_{neb}(B-V) \\ \nonumber
& \text{with} \ \ \ \ f = 0.44 \pm 0.03 \ .
\label{eBV}
\end{align}

We note that in the original calibration of eq. (3) two different reddening curves were used to measure the continuum and the nebular extinction (\citealt{Fitzpatrick99} and \citealt{Calzetti00}, respectively). If the Calzetti reddening curve is used to measure both the nebular and continuum extinction components, as done in this work, the local \emph{f}-factor becomes \emph{f}=0.58, instead of the canonical \emph{f}=0.44 \citep{Pannella14, Steidel14}. Hence, hereafter we refer to \emph{f}=0.58 as the ``local'' \emph{f}-factor.

Equation 3 implies that the ionized gas is about two times more extincted than the stars. The applicability of such relation in the high redshift Universe is still not clear, as already mentioned in Sect. 1. Several authors \citep[e.g.][]{Pannella14, Kashino13, Erb06, Reddy10, Price14, Wuyts13} found that the use of this equation implies to overestimate the intrinsic line flux, i.e. the H$\alpha$ line is less attenuated in the high redshift galaxies with respect to the local Universe.
\\To shed some light on this issue, we investigated the differential extinction between nebular lines and continuum at high-$z$ by using the ratio between the \textbf{observed} SFR$_{H\alpha}$ and SFR$_{UV}$\textbf{, thus} uncorrected for dust attenuation. 
To validate the derived dust correction we compared our measurements of SFR$_{H\alpha}$ to the SFR$_{IR+UV}$, that we assume as the ``true'' SFR estimator.

In the following section we present the formalism and the methods to derive the H$\alpha$ differential attenuation.

\subsection{Differential extinction from H$\alpha$ to UV-based SFR indicators}
\label{ffactor_eq_section}

We use the H$\alpha$ and UV luminosities, both uncorrected for dust extinction, to derive the extra correction factor associated with the nebular lines as a function of the color excess on the continuum emission. For the reader's convenience, we report in this Sect. the relevant formalism.

We assume that the total Star Formation Rate SFR$_{tot}$ is proportional to luminosity:
\begin{equation}
SFR_{tot}=const \times L_{int}(\lambda)
\label{SFRtot}
\end{equation}
where, in the case of UV-optical wavelengths, the intrinsic luminosity L$_{int}(\lambda)$ is related to the observed L$_{obs}(\lambda)$ through the attenuation A$_{\lambda}$ (eq. \ref{flux}).
Combining equations \ref{flux} and \ref{SFRtot}, we obtain for the observed SFR:
\begin{align}
 & SFR_{uncorr}=const \times L_{obs}(\lambda) = 10^{-0.4A_{\lambda}} \times SFR_{tot} \ .
\end{align}
\textbf{Considering the UV and the H$\alpha$ SFR tracers and assuming that the intrinsic SFRs in absence of dust attenuation are the same (SFR$_{H\alpha}$ = SFR$_{UV}$), we obtain that the ratio between the observed SFRs is related to the differential extinction in the H$\alpha$ emission line:}
\begin{align}
& \frac{SFR_{H\alpha ,uncorr}}{SFR_{UV,uncorr}}=\frac{10^{-0.4A_{H\alpha}} \times SFR_{tot}}{10^{-0.4A_{UV}} \times SFR_{tot}}= \nonumber \\
& =10^{-0.4E_{star}(B-V)\times [\frac{k(H\alpha)}{f}-k(UV)]} \\
& \text{or in logarithmic units} \nonumber \\
& log[\frac{SFR_{H\alpha, uncorr}}{SFR_{UV, uncorr}}]=-0.4\left[\frac{k(H\alpha)}{f}-k(UV) \right]\times E_{star}(B-V) \ .
\label{linearFit}
\end{align}
\textbf{We can then quantify the differential extinction of nebular lines through a linear fit in the plane E$_{star}(B-V), log[\frac{SFR_{H\alpha,uncorr}}{SFR_{UV,uncorr}}]$ from eq. \ref{linearFit}.}

\textbf{We stress here that in this analysis we impose that the \textbf{intrinsic} H$\alpha$ and UV luminosities are tracing the same stellar populations and this condition could be in principle not satisfied, as these two SF tracers are sensitive to different star formation timescales.
In fact, the UV luminosity is dominated by stars younger that $10^{8}$ yr while only stars with masses higher than 10 M$_{\odot}$ and lifetimes lower than $\sim$ 20 Myr contribute significantly to the ionization of the HII regions, i.e. to the H$\alpha$ luminosity \citep{Kenni98}. However, our assumption seems to be reasonable from a statistical point of view, as we would need to observe a galaxy when the formation of stars with ages $\textless 10^{7}$ yrs (i.e. the Star Formation traced by L(H$\alpha$)) is already exhausted, while the formation of stars with ages $10^{7}$ $<t<$ $ 10^{8}$ yr is already in place (i.e. sources that produce L(UV) but do not contribute significantly to L(H$\alpha$)), to have L(H$\alpha$) substantially different from L(UV)).}

\subsection{Measurements of the continuum extinction}
\label{continuum attenuation}

After we have parametrized the \emph{f}-factor and measured the ratio SFR$_{H\alpha,uncorr}$/SFR$_{UV,uncorr}$, we need to estimate the continuum color excess E$_{star}$(B-V). 

Taking advantage of the far-IR measurements available for our sample, we derived the continuum attenuation A$_{IRX}$ from the ratio L$_{IR}$ /L$_{UV}$ \citep{Nordon12}: 
\begin{equation}
A_{IRX}=2.5 log[\frac{SFR_{IR+UV}}{SFR_{UV}}] \ .
\label{Airx}
\end{equation}

We can independently infer the continuum attenuation also from the UV spectral slope $\beta$ because it is a sensitive indicator of dust attenuation. 
\textbf{The intrinsic shape of the UV continuum spectrum for a star-forming galaxy is nearly flat in F$(\lambda)$, even considering two extreme situations: an instantaneous burst of star formation and a constant star formation rate. 
For the case of the instantaneous burst, the usually assumed burst duration is typically $\sim 20$ Myr. During the first 2$\times 10^7$ yr, the stars contributing to the flux in the range $\lambda \in [1200 - 3200]$ {\AA} have had no time to evolve off the main sequence, so that the shape of the intrinsic spectrum in the UV wavelength range of interest has not changed. Considering instead a region of constant star formation, the continuous generation of new stars keeps the shape of the UV spectrum roughly constant. In conclusion, we can reasonably assume that the intrinsic shape of the UV spectra of star-forming galaxies is constant so each deviation from this intrinsic shape is produced by dust \citep{Calzetti94}.}

We then computed the continuum attenuation at 1600 {\AA} (A$_{1600}$) from $\beta$ using the calibration from \cite{Meurer99}, derived on a local sample of starburst galaxies:
\begin{equation}
A_{1600}=4.43 + 1.99 \beta \ .
\label{Abeta}
\end{equation}
Here the UV-slope parameter $\beta$ is defined as the linear interpolation of the observed spectrum (or in the lack of it from the photometric data) in the rest-frame wavelength range $\lambda \in [1250-2600]${\AA}.
The applicability of the Meurer relationship in different ranges of redshift has been demonstrated by several authors \textbf{\citep[e.g.][]{Reddy12, Buat12, Talia15}}.

In our case, since we lack observational data in this wavelength interval for the majority of the sample, we estimated $\beta$ from a linear interpolation of our best-fit MAGPHYS spectral model discussed in Sect. 4.1, as done e.g. in \cite{Oteo14}. 
To test the robustness of this slightly model-dependent estimate of the UV-slope, we also computed $\beta$ from the UV rest-frame photometry, when available (i.e. for \textbf{13} objects), and found that the two measurements are in good agreement. For more details see Appendix \ref{AppendixA}.

\begin{figure}[h!]
\centering
\resizebox{\hsize}{!}{\includegraphics{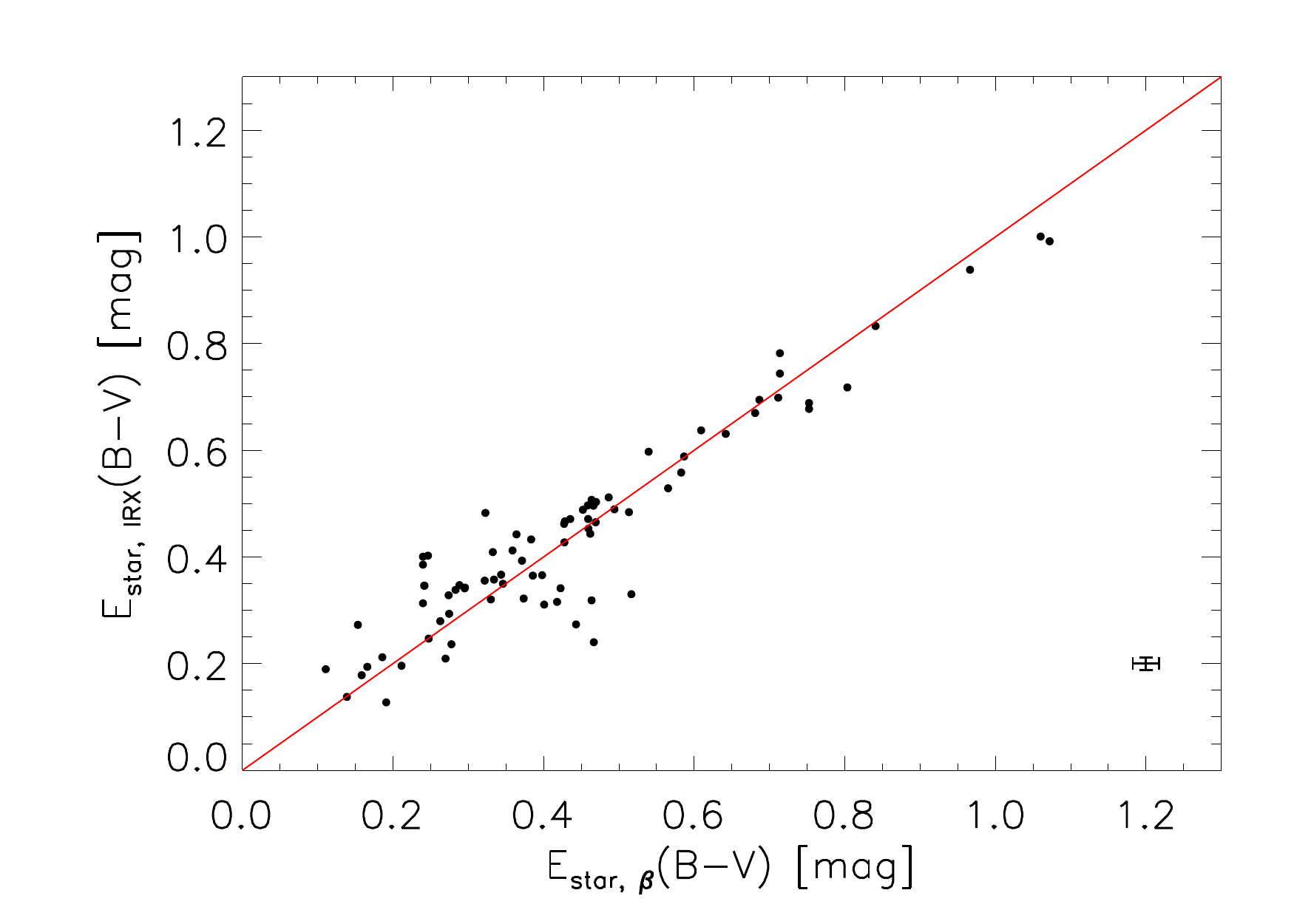}}
\caption{Comparison between the continuum extinction $E_{star}(B-V)$ derived from the UV spectral slope $\beta$ (\emph{x-axis}) and the continuum extinction derived from the infrared excess IRX (\emph{y-axis}). The red line is the 1:1 correlation relationship. In the lower part of the plot on the right are indicated the typical size of the errorbars.}
\label{EirxBeta}
\end{figure}

The attenuation is related to the color excess E$_{star}$(B-V). Assuming the reddening curve of \cite{Calzetti00} we have:
\begin{equation}
E_{star,IRX}(B-V)=\frac{A_{IRX}}{k(UV=1600)}
\label{irxequation}
\end{equation}
\begin{equation}
E_{star,\beta}(B-V)=\frac{A_{1600}}{k(UV=1600 \AA)} \ .
\end{equation}
These two quantities are consistent to each other, as displayed in Figure \ref{EirxBeta}. The correlation between E$_{star,\beta}$(B-V) and E$_{star,IRX}$(B-V) is very good, confirming the reliability of our estimate of continuum dust extinction, and has an intrinsic scatter\textbf{, i.e. not accounted for by the experimental errors \citep{AkritasBershady96},} of $\sigma_{intr} \sim 0.061$, that represent 95\% of the total scatter ($\sigma_{tot} \sim 0.064$). \textbf{Estimates of $\sigma_{intr}$ and $\sigma_{tot}$ are obtained with the IDL routine mpfit.pro \citep{mpfit}}. 

Note that, since our UV-slope $\beta$ is measured from SEDs computed with MAGPHYS, which consistently accounts for both the UV and far-IR emission, this could partially explain the high correlation degree between E$_{star,\beta}$(B-V) and E$_{star,IRX}$(B-V). In any case, this comparison is useful to verify the \textbf{accuracy} of our continuum extinction measure.

Both measurements of E$_{star}$(B-V) will be considered in the subsequent analysis to evaluate the differential extinction, i.e. the \emph{f}-factor.

The errors for E$_{star, \beta}$(B-V) were derived by propagating those on $\beta$ (see Appendix \ref{AppendixA} for the calculations of the errors on the UV slope $\beta$), while the errorbars for E$_{star, IRX}(B-V)$ are estimated via MC simulations, by varying L$_{IR}$ and L$_{UV}$ within the 1$\sigma$ errorbars.

\subsection{Extra extinction from the \emph{f}-factor}
Following the formalism of section \ref{ffactor_eq_section}, we evaluated the \emph{f}-factor as a function of the color excess on the continuum by using the ratio SFR$_{H\alpha,uncorr}$/SFR$_{UV,uncorr}$ and the two measurements of the color excess on stellar continuum, E$_{star, \beta}$(B-V) and E$_{star, IRX}$(B-V). 
To derive the \emph{f}-factor we fitted a linear function $y=s \times x$ to the data points through a minimization of a weighted $\chi^2$ \textbf{(i.e. a least square fit obtained by weighting each data point for its error, both in the \emph{x} and \emph{y} axes)}, where $y$=$log($SFR$_{H\alpha,uncorr}/$SFR$_{UV,uncorr}$) and $x=$E$_{star}$(B-V). 
To obtain this fit we used the routine mpfit.pro, and took care of checking the result by a MC simulation by inserting a scatter in the x-coordinate value for each datapoint.
The \emph{f}-factor is then finally obtained from the slope $s$ of the best-fit line as 
\begin{equation}
f = \frac{0.4 \times k(H\alpha)}{0.4 \times k(UV) - s} \ .
\label{Feq}
\end{equation}
The uncertainty for the \emph{f}-factor was estimated by propagating the error on the best-fit parameter $s$. 

Figure \ref{Betaerror} reports SFR$_{H\alpha,uncorr}$/SFR$_{UV,uncorr}$ as a function of the continuum color excess. The red solid line in Figure \ref{Betaerror} corresponds to Equation \ref{linearFit} with \emph{f}=0.93 ($\pm$ 0.065) that is the best-fit value from our data distribution. Hence we found that the \emph{f}-factor required to match SFR$_{H\alpha}$ and SFR$_{UV}$ in our sample is larger than the local value (\emph{f}=0.58, blue dashed line in Fig. \ref{Betaerror}) and turns out to be also higher than the values computed by \cite{Kashino13} on a sample with \textbf{slightly} higher $z$ and different selection criteria (\emph{f}=0.69 and 0.83, black dot-dashed and green dot-dot-dashed lines in Fig. \ref{Betaerror} respectively).

The value of the \emph{f}-factor does not change using either E$_{star, \beta}$(B-V) or E$_{star, IRX}$(B-V): this result is in line with the high correlation of these two measurements, as seen in Fig. \ref{EirxBeta}.
\textbf{
\begin{figure}[h!]
\centering
\resizebox{\hsize}{!}{\includegraphics{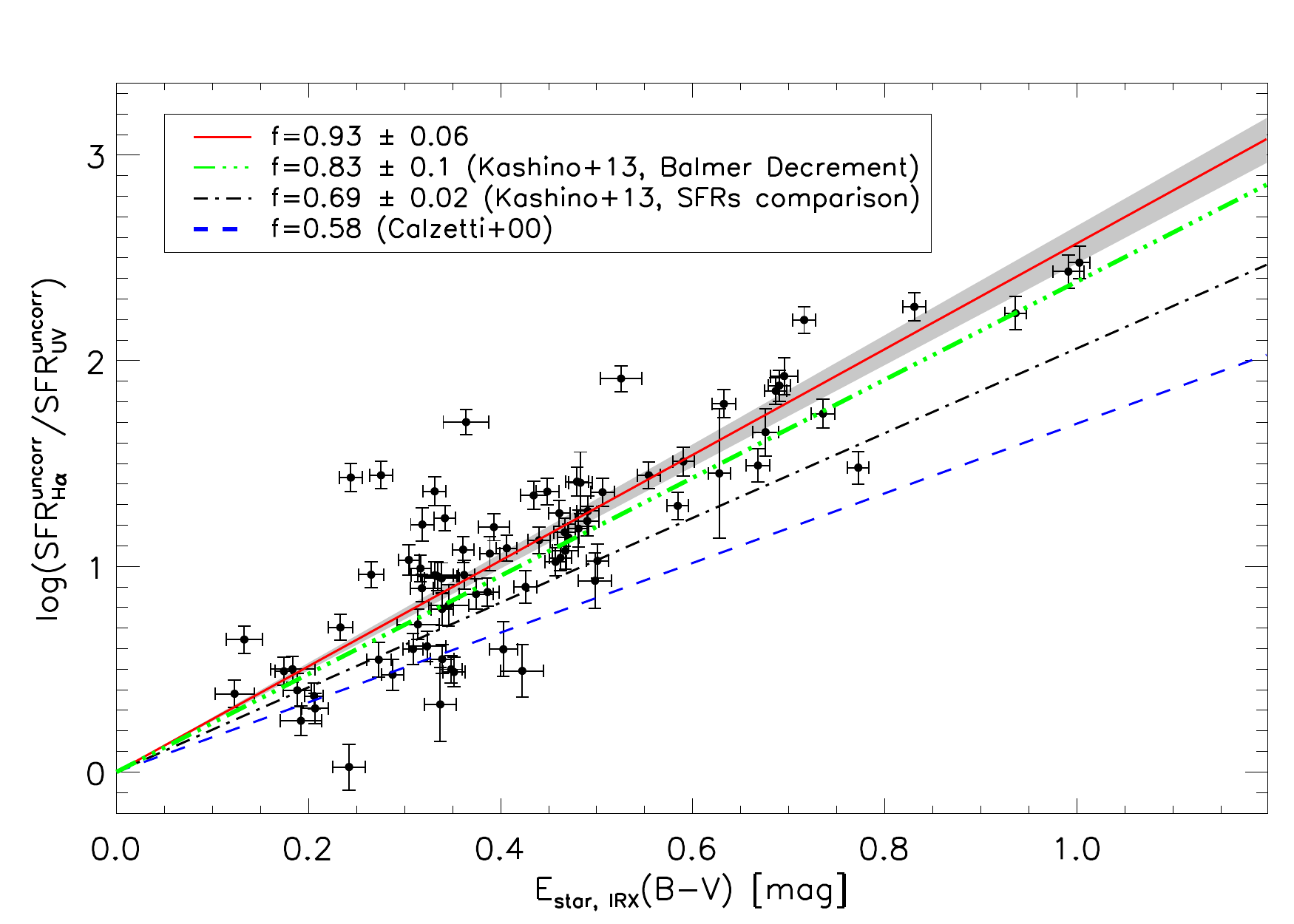}}
\caption{ \textbf{Ratio of H$\alpha$ to UV-based SFRs (not corrected for dust extinction) as a function of $E_{star}(B-V)$ derived from the IRX ratio.} The lines are the equation \ref{linearFit} with different values of \emph{f} from \cite{Kashino13} and \cite{Calzetti00}, as the legend indicates. The grey shaded area marks the confidence interval for our estimate of the \emph{f}-factors.}
\label{Betaerror}
\end{figure}
}

\subsection{Testing the dust correction: comparison between SFR$_{H\alpha}$ and SFR$_{IR+UV}$}
\label{compareSFRiruvSection}
\begin{figure*}[h!]
\centering
\resizebox{\hsize}{!}{\includegraphics[angle=90]{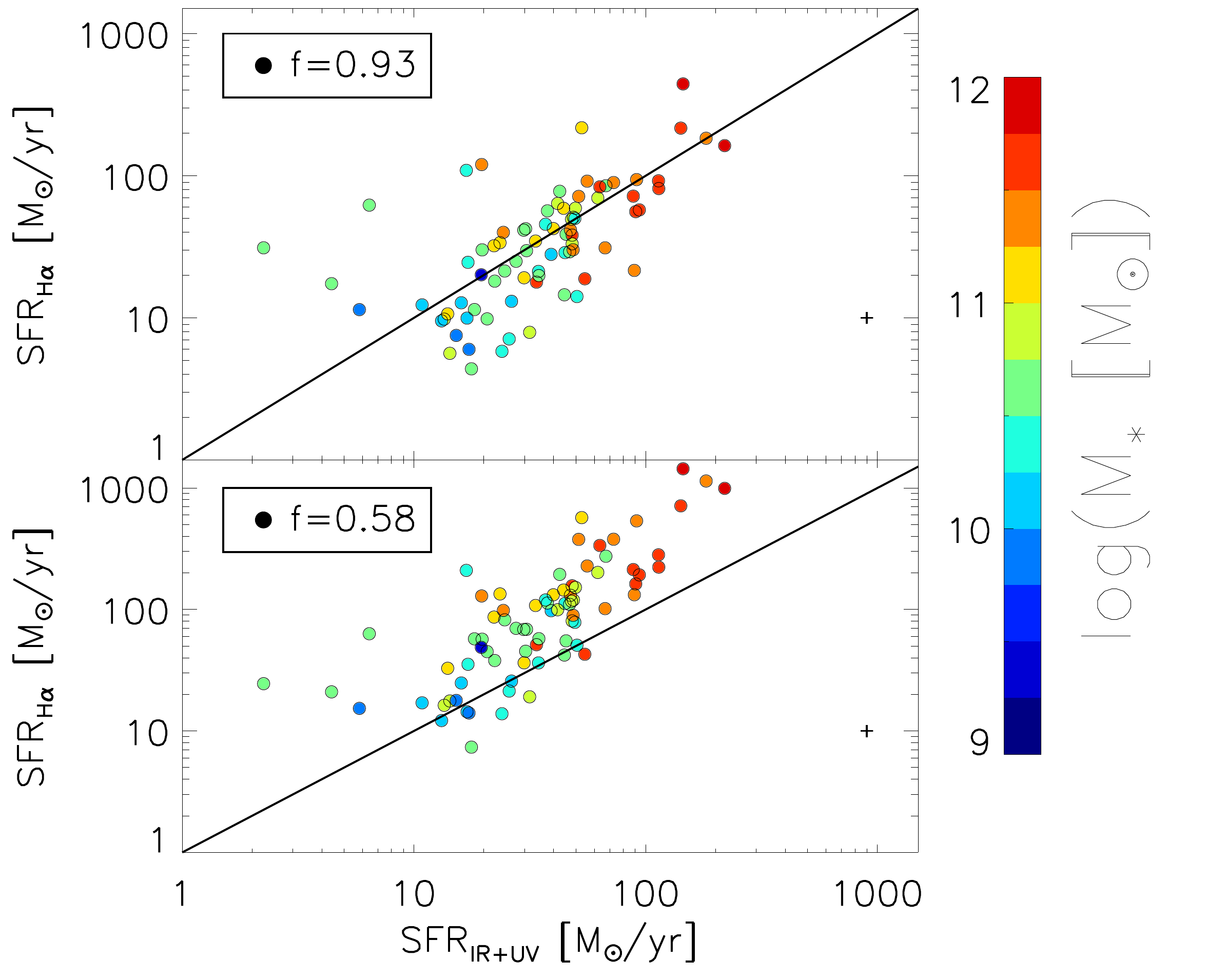}}
\caption{Comparison between SFR$_{IR+UV}$ and SFR$_{H\alpha}$ when varying the dust correction for the H$\alpha$ emission. The lower panel shows the case in which L$_{H\alpha}$ are corrected for dust attenuation by using the classical prescription of Calzetti (\emph{f=0.58}) while in the upper panel the H$\alpha$ luminosities are corrected with our dust correction (\emph{f=0.93}). 
The black solid line is the 1:1 correlation line and the different colors mark the stellar masses, as indicated in the vertical color-bar. In the lowest part of each panel are also shown the median errors on the SFR measurements.}
\label{compareSFR}
\end{figure*}

We tested the reliability of our extinction correction in Figure \ref{compareSFR}, where we compare the SFR$_{IR+UV}$ with the SFR$_{H\alpha}$, corrected for dust extinction using both the prescription derived in this work \textbf{(i.e. E$_{star, IRX}$(B-V) and \emph{f}=0.93, upper panel)}, and that derived by \cite{Calzetti00} for local galaxies \textbf{(E$_{star, IRX}$(B-V) and \emph{f}=0.58, lower panel)}. The results show that a better agreement between SFR$_{H\alpha}$ and SFR$_{IR+UV}$ is obtained by applying our correction factor, with a median ratio SFR$_{H\alpha}$/SFR$_{IR+UV}$=0.88 and a median relative scatter $\textbar$SFR$_{IR+UV}$-SFR$_{H\alpha}\textbar$/(1+SFR$_{IR+UV}$)=0.38. On the contrary, the Calzetti et al. prescription would lead to overestimate the SFR$_{H\alpha}$  by a factor up to 3.3 above SFR$_{IR+UV}\textgreater$50 M$_{\odot}$/yr.
We emphasize that our recipe for the nebular dust attenuation seems to ``fail'' for the tail of objects with SFR$_{IR+UV} \in [10 - 20]$M$_{\odot}$/yr and M$_{\star} \sim 10^{10}$M$_{\odot}$ (light blue points in Fig. \ref{compareSFR}). For these sources our dust correction underestimates the SFR$_{H\alpha}$ with respect to SFR$_{IR+UV}$, while using the Calzetti prescription the two SFRs are in slightly better agreement.
This could imply that the \emph{f}-factor is an increasing function of SFR and M$_{\star}$. 

However, the disagreement between SFR indicators at SFR$_{IR+UV} \lesssim 20 $ M$_{\odot}$/yr can be due also to an overestimate of SFR$_{IR+UV}$ rather than a problem of the dust correction. At lower M$_{\star}$ and SFRs in fact, the contamination to the heating of dust by an older stellar population \citep[the so-called ``cirrus'' component, e.g.][]{Kenni09} became not negligible and can cause to overestimate L$_{IR}$.
Finally, the trend seen in Figure \ref{compareSFR} could also be a consequence of the selection criterion used in this work, considering that SFR$_{IR+UV} \sim 15$M$_{\odot}$/yr is the value that corresponds to the 3$\sigma$ flux limit at 160 $\mu$m at $z \sim 1$ as estimated using the median SEDs of \cite{Magdis12} (see also Fig. \ref{parent_sample} of section \ref{final_sample}).

\subsection{A$_{H\alpha}$ vs M$_{\star}$}

\begin{figure}[h!]
\centering
\resizebox{\hsize}{!}{\includegraphics{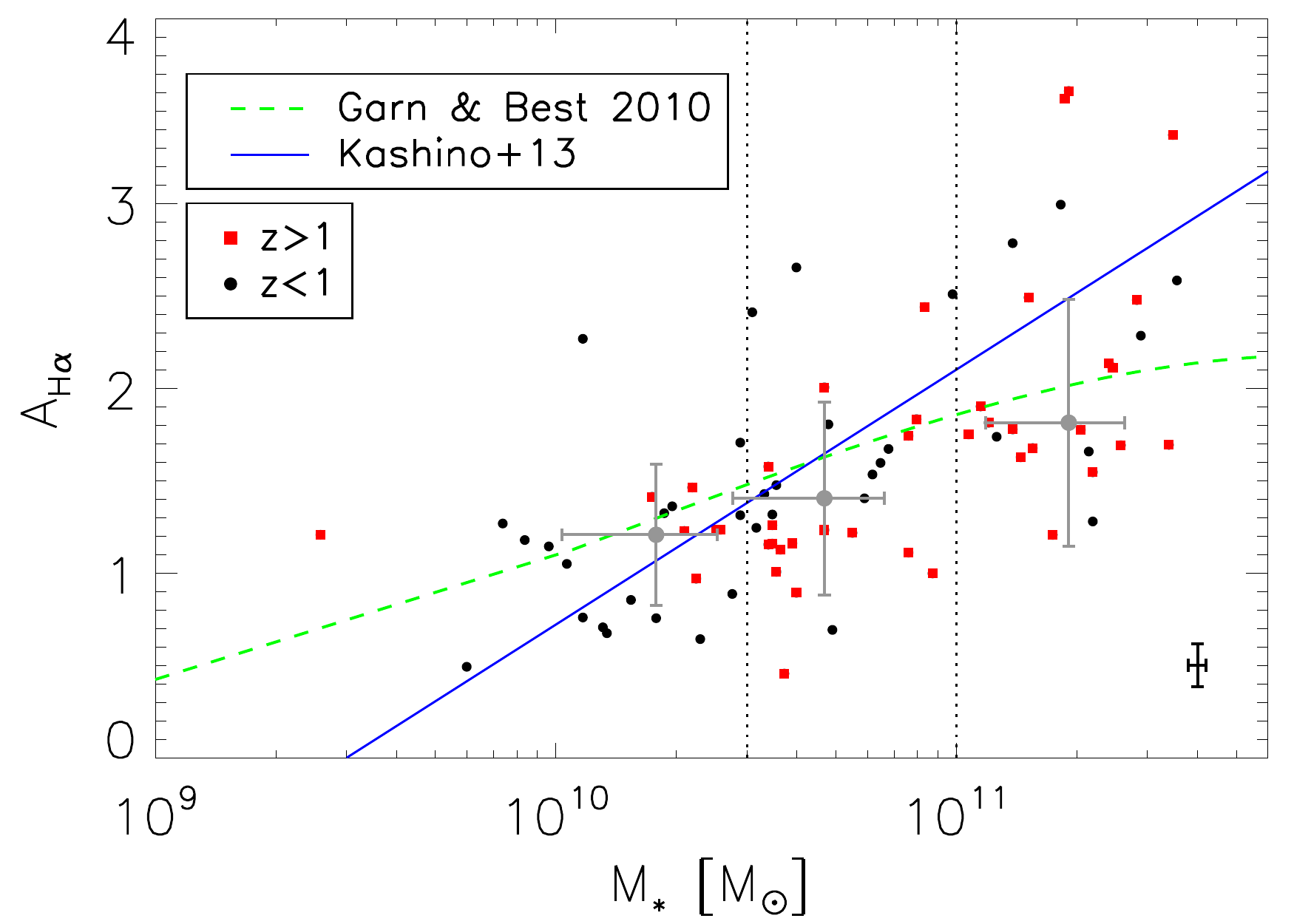}}
\caption{A$_{H\alpha}$ as a function of stellar masses. The black filled circles are the sources at $z \textless 1$ while red filled squares are objects with $z \textgreater 1$. The attenuation is computed from \textbf{E$_{star, IRX}$(B-V)} by using \emph{f=0.93}. The grey filled circles are the median values of A$_{H\alpha}$ in three mass bins (M$_{\star}$ $\textless$ $3 \times 10^{10}$ M$_{\odot}$, $3 \times 10^{10}$ $\leq$ M$_{\star}$ $\textless$ $1.7 \times 10^{11}$ M$_{\odot}$, M$_{\star}$ $\geq$ $1.7 \times 10^{11}$ M$_{\odot}$) and the errorbars are the median scatter for A$_{H\alpha}$ and M$_{\star}$ in each mass bin. The two vertical dashed lines highlight the mass bins (M$_{\star}$ $\textless$ $3 \times 10^{10}$ M$_{\odot}$, $3 \times 10^{10}$ $\leq$ M$_{\star}$ $\textless$ $1.7 \times 10^{11}$ M$_{\odot}$, M$_{\star}$ $\geq$ $1.7 \times 10^{11}$ M$_{\odot}$).}
\label{AhaMstar}
\end{figure}

Figure \ref{AhaMstar} shows the relationship between the attenuation A$_{H\alpha}$ and stellar masses. Here we converted the continuum attenuation \textbf{E$_{star, IRX}(B-V)$} to A$_{H\alpha}$ by using \emph{f=0.93} and the k($\lambda$) from \cite{Calzetti00}. Despite the high dispersion of the data, we can observe an excess of A$_{H\alpha}$ at the highest masses with respect to the local relationship of \cite{GarnBest} (green dashed line in Figure \ref{AhaMstar}). We report also the relationship of \cite{Kashino13}, derived at $z$ $\sim$ $1.6$ (blue solid line). The grey filled dots of Figure \ref{AhaMstar} represent the median values of A$_{H\alpha}$ obtained in three M$_{\star}$ bins as reported in the figure.
The errorbars are the median scatter of A$_{H\alpha}$ and M$_{\star}$ in each bin. 
Within the large scatter in the data, the median values in each bin are consistent with both the local and the higher-$z$ relationship, in agreement with the results of \cite{Ibar13}.
Of course, there may be selection effects in operation in the graph disfavoring the detection of galaxies with 
high values of A$_{H\alpha}$, that may be more important for the lower mass objects with intrinsically faint H$\alpha$ flux. Our combined selection requiring detection of the H$\alpha$ line may be somewhat exposed to such an effect.
\textbf{To confirm if an evolution with $z$ exists in the dust properties of star forming galaxies it would be necessary to have deeper observations for the most attenuated galaxies.} This aspect of the analysis will be investigated in a further paper. 

\subsection{Main sequence at $z$ $\sim$ 1}
\begin{figure}[h!]
\centering
\resizebox{\hsize}{!}{\includegraphics[angle=90]{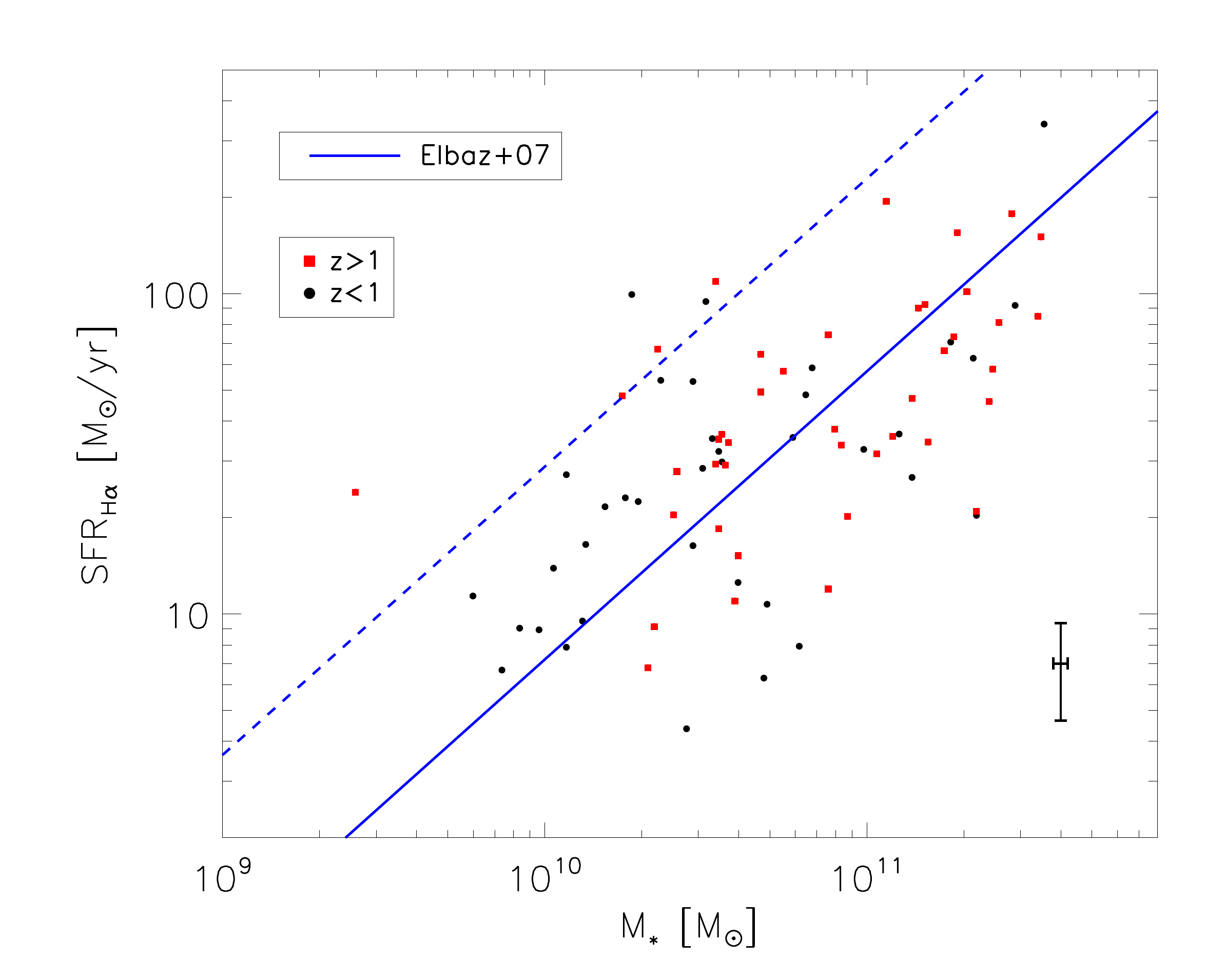}}
\caption{SFR$_{H\alpha}$ versus $M_{\star}$ with L$_{H\alpha}$ corrected by dust with our recipe. The blue solid line reports the MS relationship from \cite{Elbaz07} at $z=1$ and the blue dashed line is the corresponding $4 \times MS$. The black filled circles are the sources at $z \textless 1$ while the red filled squares are the objects at $z \textgreater 1$. The median errors for M$_{\star}$ and SFR$_{H\alpha}$ are reported in the lower part of the plot on the right.}
\label{MSfromHa}
\end{figure}

\textbf{In Figure \ref{MSfromHa} we report an updated version of the relation between the stellar mass and the SFR, already presented in Figure \ref{parent_sample}, where SFR is now computed from the H$\alpha$ luminosities, corrected for dust extinction according to our reference dust correction. As already mentioned in Section \ref{final_sample}, our sources do not span the overall range covered by the star-forming Main Sequence population at $z \sim 1$, since our far-IR selection favor the detection of sources with SFR$\textgreater$10-20 M$_{\odot}$/yr. However, we note that above stellar masses on the order $3\times10^{10}$M$_{\odot}$, our sample is basically representative of a mass-selected sample that traces the underlying MS at the same redshift.}

\section{Caveats: the many faces of the \emph{f}-factor}
\label{caveats}

Values of the \emph{f}-factor estimated in the literature range from  0.44 to $\sim 1$ \textbf{, as shown in Table \ref{f_literature},} and in this section we briefly discuss the likely origins of these differences.  An \emph{f}-factor less than unity is currently interpreted as implying more average obscuration affecting the line emitting regions (HII regions) compared to the average obscuration affecting the hot stars emitting the UV continuum, i.e., \emph{f} $\ne 1$ would be the result of a geometrical effect with the spacial distribution of line emitting regions being different from that of continuum emitting stars.  Actually, the reality may be somewhat more complicated.

First of all, the \emph{f}-factor can be derived in two radically different ways: either by comparing two extinctions or two SFRs. In the former case the H$\alpha$ extinction derived from the Balmer decrement is compared to the extinction in the UV as derived from either the UV slope or from Equation (10). Then one has to adopt one specific reddening law $k(\lambda)$. In the latter case, the \emph{f}-factor is estimated  by enforcing equality between the SFR derived from the H$\alpha$ flux with the SFR derived from another indicator, such as SFR(UV), SFR(UV+IR) or SFR from SED fitting, etc. 

When the \emph{f}-factor is derived by comparing H$\alpha$ and UV extinction, then the result depends on the adopted reddening law $k(\lambda)$, hence it reflects at once both the mentioned geometrical effect and possible departures from the adopted reddening law. As well known, the reddening law is not universal, not even within the Local Group. These two contributions to determine the value of \emph{f} can barely be disentangled.
When the \emph{f}-factor is derived by forcing agreement between SFR(H$\alpha$) and the SFR from another indicator, then it becomes a fudge factor to compensate for the relative biases of the two SFR estimators, neither of which will be perfect.

Moreover, the measured H$\alpha$ flux is subject to dust absorption in two distinct ways: first, each H$\alpha$ photon has a probability $10^{-0.4A_{H\alpha}}$ to escape the galaxy, but, second, dust absorption in the Lyman continuum reduces the number of produced H$\alpha$ photons by a factor $10^{-0.4A_{Lyman-cont.}}$, where $A_{Lyman-cont.}$ is the extinction in the Lyman continuum. 
With few exceptions (e.g. \citealt{Boselli09}), this second aspect is generally ignored, in the hope that an empirical calibration of SFRs may subsume in it also this part of the involved physics. In any event, the resulting \emph{f}-factor depends on the actual extinction law of each galaxy, extending from the optical all the way to the Lyman continuum, on the geometry of the emitting regions, as well as on the relative systematic biases in the relations connecting SFRs to observables.

Besides these aspects, derived \emph{f} values can also depend on the specific galaxy sample from which it is derived.
For example, in our approach  the far-IR selection criterion leads to select objects with strong levels of dust obscuration in the UV. The other requirement is to select objects with a strong H$\alpha$ emission (F$_{obs}(H\alpha) \textgreater 2.87 \times 10^{-17}$ $erg/s$ and S/N higher than $\sim$ 3), hence with lower levels of H$\alpha$ attenuation. The two selection criteria partially conflict with each other and combined favor sources with E$_{neb}$(B-V) $\sim$ E$_{star}$(B-V).
To understand how the selection criterion influences the results we can compare our analysis with the work of \cite{Kashino13}. \textbf{Our analysis was performed following the same approach of \cite{Kashino13} but our sample has different selection criterion and size (168 sBzK galaxies for Kashino, 79 far-IR sources in this work) and we obtained an \emph{f}=0.93, that is  35\% larger than the Kashino result.} In the case of rest-frame UV-selected galaxies, such as in \cite{Erb06}, this bias favors galaxies with low extinction which may result in a different $f$ value compared to the case of samples including also highly reddened galaxies.

All of these considerations imply that different indicators lead to different estimates of the \emph{f}-factor. For example, if we consider the ratio SFR$_{H\alpha}^{unc}$/SFR$_{IR+UV}$, we get (see e.g. eq. \ref{linearFit})
\begin{equation}
\frac{SFR_{H\alpha ,uncorr}}{SFR_{IR+UV}}=10^{-0.4 A_{H\alpha}} = 10^{-0.4 \frac{E_{star}(B-V)}{f} k(H\alpha)}. 
\end{equation}
Our data would imply in this case \emph{f}$\simeq 0.85$, however not very significantly different from our best-guess of \emph{f}$\simeq 0.93$.

Also the estimate of E$_{star}(B-V) $ and its errors influences the estimate of the \emph{f}-factor, leading to results that can vary from \emph{f} $\sim$ 0.4 to a value larger than 1.

The last point to consider is again the reddening law $k(\lambda)$ assumed in the analysis. In the literature one finds  very different trends for $k(\lambda)$, such as the presence of a bump at $\sim 2200${\AA} \citep[for the LMC]{Fitzpatrick99} or a smoother trend \citep[for a starburst galaxy]{Calzetti00}, \textbf{and also very different values for its normalization R$_{V}\equiv$ A$_{V}/$E$_{star}$(B-V), that vary from 3.1 \citep{Cardelli89} to 4.05 \citep{Calzetti00}}. At redshift $\sim 2$ both galaxies with and without the 2200 \AA\ bump appear to coexist \citep{Noll09}. The shape and the normalization of the assumed $k(\lambda)$ strongly influences the value of \emph{f}, both for direct or indirect measurements. As an exercise we derived the \emph{f}-factor using different $k(\lambda)$ expressions in eq. \ref{Feq}. Table \ref{exercise} summarizes the results, showing that the value of the \emph{f}-factors ranges from $\sim$ 0.7 to $\sim 1.2$. 

\textbf{In summary, the \emph{f}-factor may offer a fair measure of the relative extinction of emission lines and the stellar continuum, still however relying on not-completely well defined and understood ingredients.
}

\begin{center}
\begin{table}[h!!!]
\caption{Summary of the values obtained for the \emph{f}-factor as a function of the assumed reddening curve $k(\lambda)$: the first column is the $k(\lambda)$ assumed to compute the UV attenuation at 1600 {\AA}, the second column is the $k(\lambda)$ used to compute the H$\alpha$ attenuation while the last column is the \emph{f}-factor, obtained from eq. \ref{Feq}.}
\label{exercise}
\centering
\begin{tabular}{ccc}
\hline \hline
$k(UV)$ & $k(H\alpha)$ &\emph{f}-factor \\ \hline
\cite{Calzetti00} & \cite{Calzetti00} & 0.93 \\
\cite{Calzetti00} & \cite{Fitzpatrick99} & 0.66 \\
\cite{Reddy15} & \cite{Reddy15} & 0.96 \\
\cite{Reddy15} & \cite{Fitzpatrick99} & 1.19 \\ \hline \hline
\end{tabular}
\end{table}
\end{center}

\section{Summary and conclusions}
\label{Conclusions}

In this work we analyzed the near-IR spectra of 79 star forming galaxies at $z \in [0.7-1.5]$, acquired from the 3D-HST survey . The sources were selected in the far-IR from the \textit{Herschel}/PACS observations: the PACS catalogs were associated with the 3D-HST observations using the IRAC positions of the PACS sources. From the near-IR spectra we measured the H$\alpha$ fluxes and the spectroscopic redshifts of the whole sample.

We computed the SEDs with the MAGPHYS software, using data from near-UV to far-IR including the GALEX-NUV, the GOODS-MUSIC optical to mid-IR catalog, the IRS-16 $\mu$m and the far-IR photometry from \textit{Herschel} PACS and SPIRE (i.e. at 70, 100, 160, 250, 350 and 500 {$\mu$}m).
From the SEDs we derived the stellar masses, M$_{\star}$, the bolometric infrared luminosities, L$_{IR}$, the UV luminosities, L$_{UV}$, and the UV slope, $\beta$.

We then evaluated the color excess E$_{star}$(B-V) from the IRX=L$_{IR}$/L$_{UV}$ ratio and from the UV slope $\beta$ and found that these two quantities are in good agreement. In our sample the color excess on the stellar continuum ranges from E$_{star}$(B-V) $\sim$ 0.1 \textit{mag} to E$_{star}$(B-V) $\sim$ 1.1 \textit{mag}.

We computed the dust attenuation on the H$\alpha$ emission E$_{neb}$(B-V) as a function of E$_{star}$(B-V) by comparing the SFR$_{H\alpha}$ and the SFR$_{UV}$, both uncorrected for extinction.
We obtained that the \emph{f}-factor, which parametrizes the differential extinction on the nebular lines, is \emph{f}=E$_{star}$(B-V)/E$_{neb}$(B-V)=0.93 $\pm$ 0.06. 
This result is consistent within the errorbars with the analysis of \cite{Kashino13} from the Balmer Decrement and of \cite{Pannella14}, performed in a similar redshift range. Our analysis is also consistent with the results of \cite{Erb06} and \cite{Reddy10} performed at higher $z$, as summarized in Table \ref{f_literature}, which collect a list of results from others works.
The good agreement found in our sample between the SFR$_{IR+UV}$ and the SFR$_{H\alpha}$ corrected for extinction using our recipe further confirm our results.

From our dust correction we then computed the attenuation A$_{H\alpha}$ as a function of M$_{\star}$. We found that A$_{H\alpha}$ is increasing with M$_{\star}$ and this trend seems to diverge from the local relationship: our sources shows an excess of A$_{H\alpha}$ with respect to the relationship of \cite{GarnBest} for M$_{\star}$ $\gtrsim$ 10$^{11}$ M$_{\odot}$, suggesting an evolution in the dust properties of star-forming galaxies with $z$.

In conclusion we found that the level of differential extinction required to match the SFR$_{H\alpha}$ with the SFR$_{IR+UV}$ is lower than in the local Universe, thus A$_{H\alpha}$ $\sim$ A$_{UV}$ for the sources in our sample. The value of the \emph{f}-factor seems to be related to the physical properties of the sample rather than be dependent on $z$. The trends of Figures \ref{compareSFR} and \ref{AhaMstar} suggest that the H$\alpha$ extinction (thus the \emph{f}-factor) is a function of SFR and M$_{\star}$. In particular we notice that in Figure \ref{compareSFR} our dust correction underestimate SFR$_{H\alpha}$ with respect to SFR$_{IR+UV}$ for sources with SFR $\lesssim$ 20 M$_{\odot}$/yr: these sources require a lower value of \emph{f}, similar to the local \emph{f}=0.58.
This trend could be explained by the two components model of dust sketched in Figure 5 of \cite{Price14}.  
A galaxy with high sSFR is supposed to have an high number of OB stars, which are located inside the optically thick birth cloud: in this case these massive stars dominate both the UV-continuum and the H$\alpha$ emissions, so the level of attenuation for the continuum and the nebular emission will be similar (A$_{UV}$ $\sim$ A$_{H\alpha}$). On the other and, for a galaxy with low sSFR the number of OB stars will be lower, so in this case the optical-UV continuum is mainly produced by the less massive stars that are located both in the birth cloud and in the diffuse ISM: in this case A$_{H\alpha}$ $\textgreater$ A$_{UV}$ since the H$\alpha$ emission is produced in a different and more dust-dense region with respect to the continuum.

In this paper we do not examine in depth the consequences of this modellistic approach for the reasons discussed in the previous section.
We defer to a future paper a more detailed analysis of the extinction properties of star-forming galaxies based on the ``Intensive Program'' (S12B-045, PI J. Silverman) with the FMOS spectrograph at the Subaru Telescope in the COSMOS field \citep{Silverman14}.

\begin{center}
\begin{table*}[h!!!]
\caption{Different values of \emph{f} obtained from literature. In the table are also specified the redshift ranges (third column of Tab. \ref{f_literature}), the types of sample and the methods implied to measure the differential extinction on nebular lines (fourth and fifth columns of Tab. \ref{f_literature}, respectively).}
\label{f_literature}
\centering
\begin{tabular}{ccccc}
\hline \hline
\textit{\textbf{Author}} & \textbf{\emph{f}} & \textbf{$z$ \textit{range}} & \textbf{\textit{Sample}} & \textbf{\textit{Method}} \\ \hline
\cite{Calzetti00} & 0.44 (0.58) & 0.003-0.05 & starburst galaxies & Balmer decrement \\ 
\cite{Kashino13} & 0.83\footnotesize{$\pm$0.10} & 1.4-1.7 & sBzk galaxies & Balmer decrement (stacked spectra) \\
" & 0.69\footnotesize{$\pm$0.02} & " & " & H$\alpha$ to UV SFRs \\
\cite{Wuyts11} & 0.44 & 0-3 & K$_s$ selected galaxies & SFR indicators \\
\cite{Price14} & 0.55\footnotesize{$\pm$0.16} & 1.36-1.5 & MS star forming galaxies & Balmer decrement (stacked spectra) \\
\cite{Pannella14} & 0.58 & $\textless$ 1 & UVJ selected galaxies & comparison between A$_{UV}$-M$_{\star}$/A$_{H\alpha}$-M$_{\star}$ \\
" & 0.77 & 1 & " & " \\
" & 1 & $\textgreater$ 1 & " & " \\
\cite{Valentino15} & 0.74\footnotesize{$\pm$0.05} & 2 & CL J1449+0856 & Balmer decrement (stacked spectra) \\
\cite{Erb06} & 1 & 2 & rest-frame UV selected galaxies & matching UV and H$\alpha$ SFRs \\
\cite{Reddy10} & 1 & 1.5-2.6 & Lyman Break Galaxies & matching X-ray, UV and H$\alpha$ SFRs \\
This work & 0.93\footnotesize{$\pm$0.06} & 0.7-1.5 & far-IR selected galaxies & H$\alpha$ to UV SFRs \\ \hline \hline
\end{tabular}
\end{table*}
\end{center}

\begin{acknowledgements} 
We thank the anonymous referee for their careful reading of the manuscript an their valuable comments.
AP, GR and AF acknowledge support from the Italian Space Agency (ASI) (Herschel Science Contract I/005/07/0).
We are grateful to Antonio Cava for the improvement of the IDL code used for the spectral analysis and Robert Kennicutt and Naveen Reddy for their comments.
We also thank Mattia Negrello for his help with the error analysis and his comments.
This work is based on observations taken by the 3D-HST Treasury Program (GO 12177 and 12328) with the NASA/ESA HST, which is operated by the Association of Universities for Research in Astronomy, Inc., under NASA contract NAS5-26555.
\end{acknowledgements}

\bibliographystyle{aa}
\bibliography{bibliography}

\appendix
\section{Computation of the UV spectral slope}
\label{AppendixA}
\subsection{$\beta$ from the best-fit SED}
\begin{figure}[h!]
\centering
\resizebox{\hsize}{!}{\includegraphics{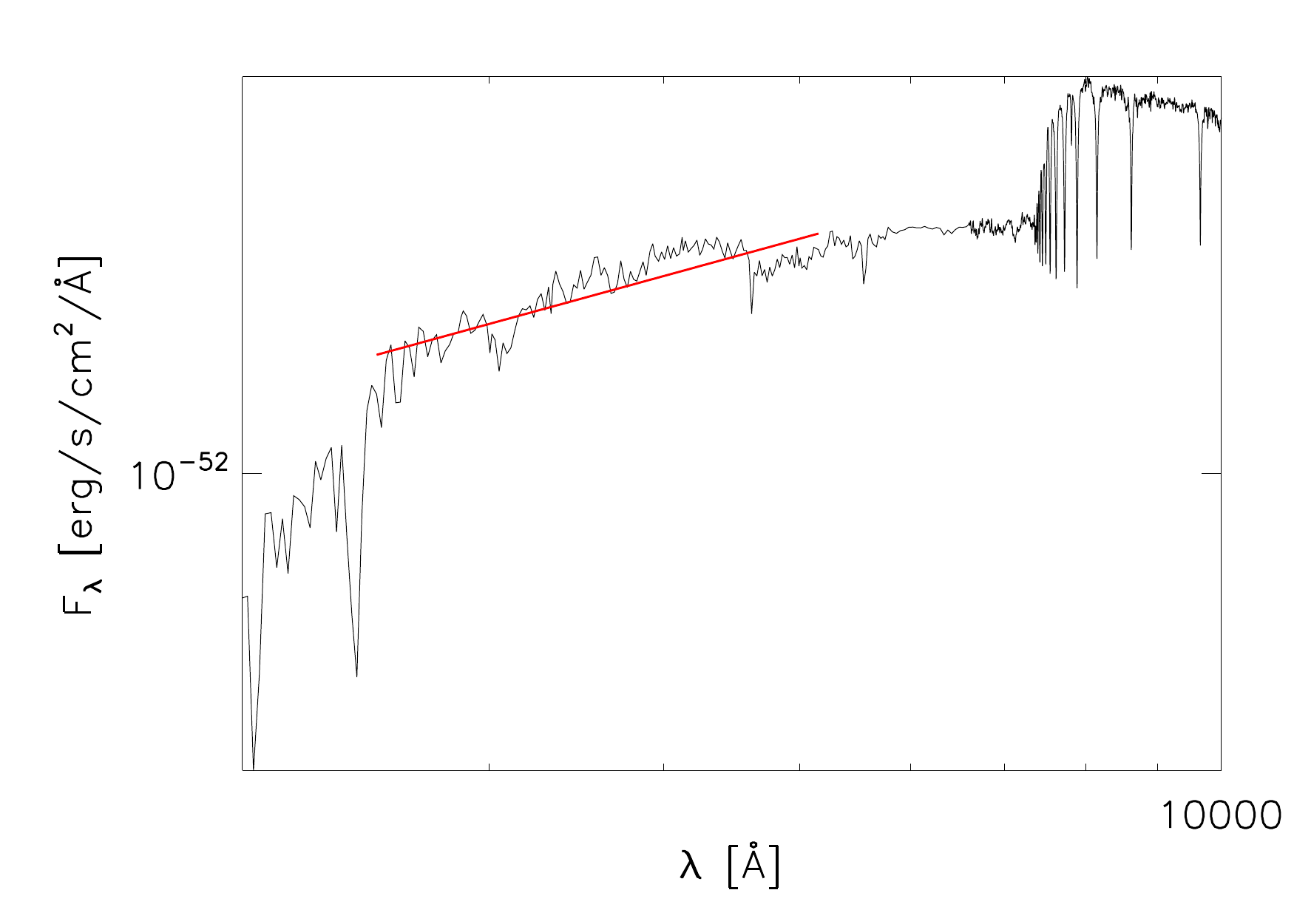}}
\caption{Linear fit (red line) to the best-fit SED in the plane $log(\lambda),log(F_{\lambda})$. }
\label{betaSED}
\end{figure}

The UV slope is derived from a linear fit to the model best-fit SED in the wavelength range $\lambda \in [1250-2600]$ \AA, lacking observations in this rest-frame range for the majority of the sample. An example of the fit is shown in figure \ref{betaSED}.
The errors related to the \textbf{$\beta_{model}$, i.e. the UV slope derived from the MAGPHYS SED,} are computed from the linear fit.
\subsection{$\beta$ from the observed photometry}
\begin{figure}[h!]
\centering
\resizebox{\hsize}{!}{\includegraphics{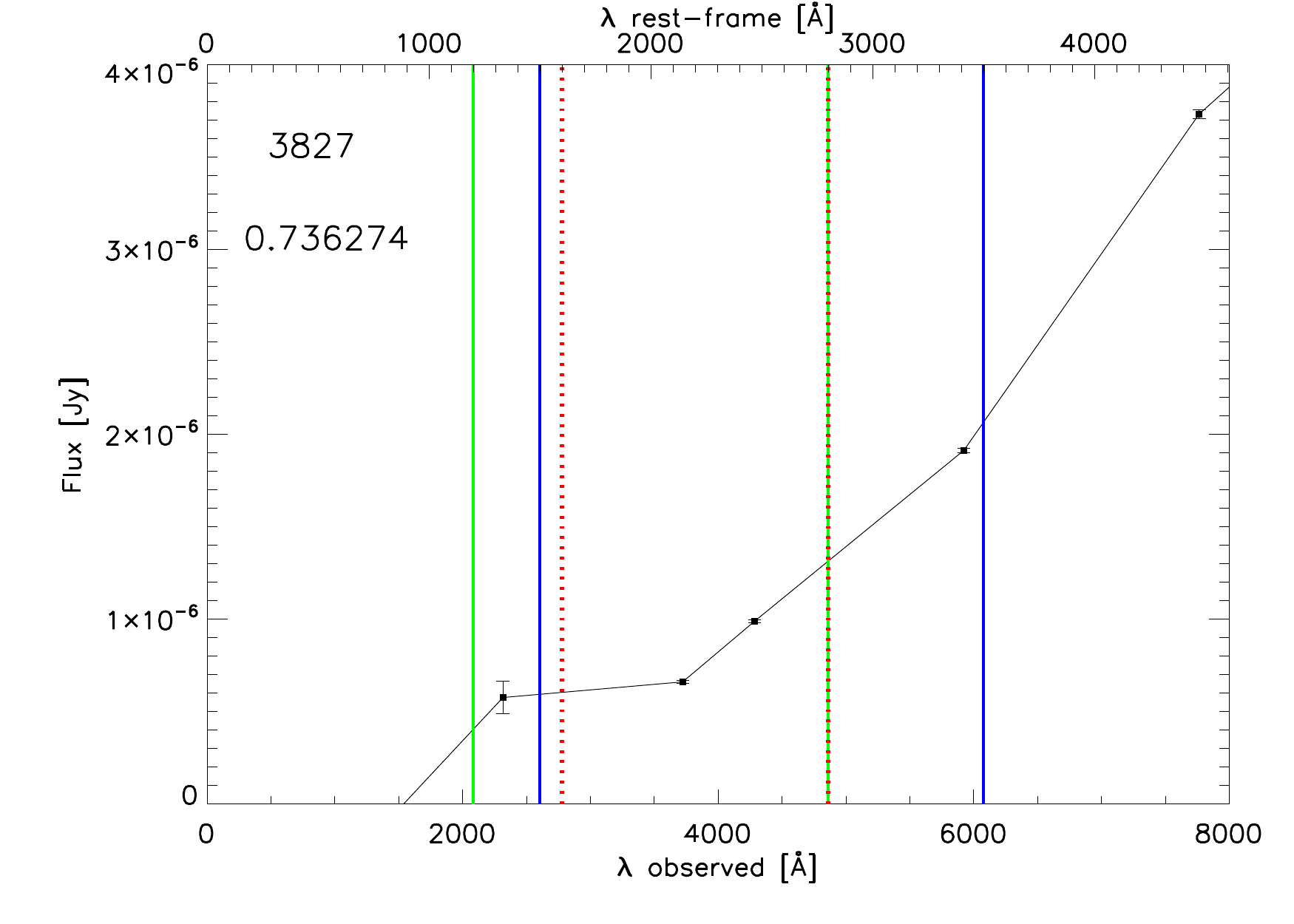}}
\caption{The plot displays the method for the computation of $\beta_{phot}$ for the source 3213 (ID MUSIC). In the figure it is specified also the redshift of the source. The black open diamonds are the observed photometry, connected with a linear interpolation (black solid line). The green vertical lines highlight the rest-frame spectral range considered to compute M$_{1}$ while the blue vertical lines mark the range for the computation of M$_{2}$. The two dash-dot red lines mark respectively the positions of 1600 {\AA} and 2800 {\AA} rest-frame.}
\label{nordonCoverage}
\end{figure}

\begin{figure}[h!]
\centering
\resizebox{\hsize}{!}{\includegraphics{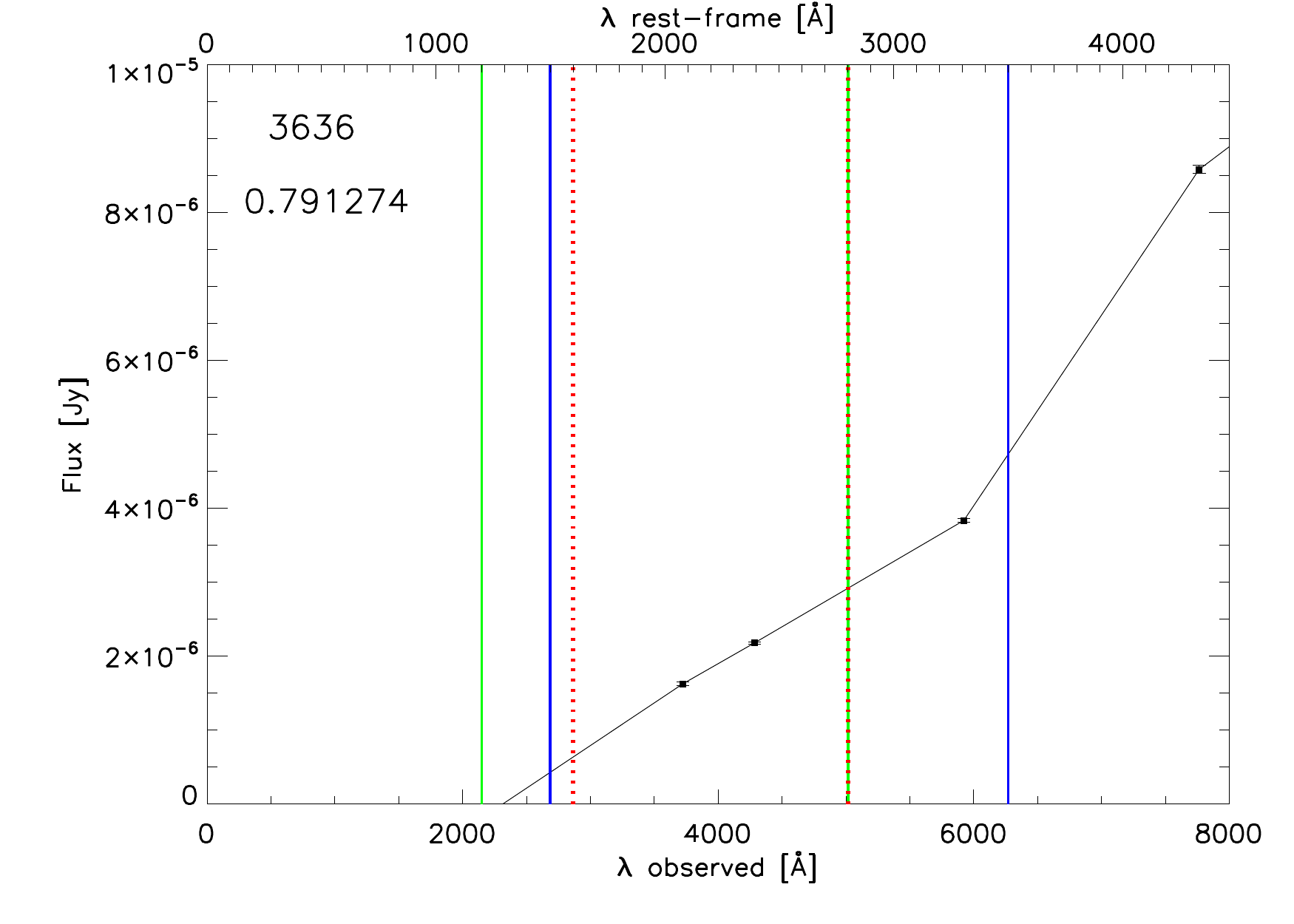}}
\caption{The plot is analogous to the previous. In this case we do not have the photometric coverage to derive $\beta_{phot}$.}
\label{nordonNoCoverage}
\end{figure}

\textbf{In order to test the validity of the estimate of \textbf{$\beta_{model}$} we computed the UV-slope also from the observed data, when is available the photometric coverage in the rest-frame range of interest}. We interpolated the observed photometry in the rest-frame range $\lambda \in [1200-3500]$ {\AA}, following the method described in \cite{Nordon13}. 
The photometric UV-spectral slope is defined as:
\begin{equation}
\beta_{phot}=\frac{-0.4(M_1 - M_2)}{log(\lambda_1/\lambda_2)}-2
\end{equation}
where M$_1$ and M$_2$ are the AB magnitudes at wavelengths $\lambda_1$=1600 {\AA} and $\lambda_2$=2800 {\AA}.
\\The rest frame $1600$ {\AA} and $2800$ {\AA} magnitudes (M$_{1600}$ and M$_{2800}$) were estimated by interpolating between the available photometric bands. To derive M$_{1600}$ we used the filters between rest-frame $\lambda \in$ $[1200-2800]$ {\AA} and interpolated the flux at $1600${\AA} (converted to AB magnitudes) by fitting a \textbf{linear function between the observed photometric bands}. To derive M$_{2800}$ we selected the filters that observe the rest-frame $\lambda \in$ $[1500-3500]$ {\AA}.
Figure \ref{nordonCoverage} shows an example for the computation of M$_{1600}$ and M$_{2800}$ for a source that have the required photometric coverage. Figure \ref{nordonNoCoverage} shows the case in which we haven't the coverage in the rest-frame. The $\beta_{phot}$ was computed for \textbf{13} sources. 
We computed the errors for $\beta_{phot}$ by using a set of MC simulations: we randomly varied the observed photometry within the errorbars and we then computed the interpolation for each set of simulated values of the observed photometry. The 1$\sigma$ uncertainty on $\beta_{phot}$ was then estimated from the width of the probability distribution function of the simulated values, assuming that this distribution has a Gaussian shape. \textbf{The median error on $\beta_{phot}$ was added in quadrature to the error of $\beta_{model}$ thus the error for the UV-slope is $\sigma^{2}_{\beta ,i}= \sigma^{2}_{\beta_{model},i} + median(\sigma^{2}_{\beta_{phot},i})$}.

\subsection{$\beta_{phot}$ vs \textbf{$\beta_{model}$}}
\begin{figure}[h!]
\centering
\resizebox{\hsize}{!}{\includegraphics{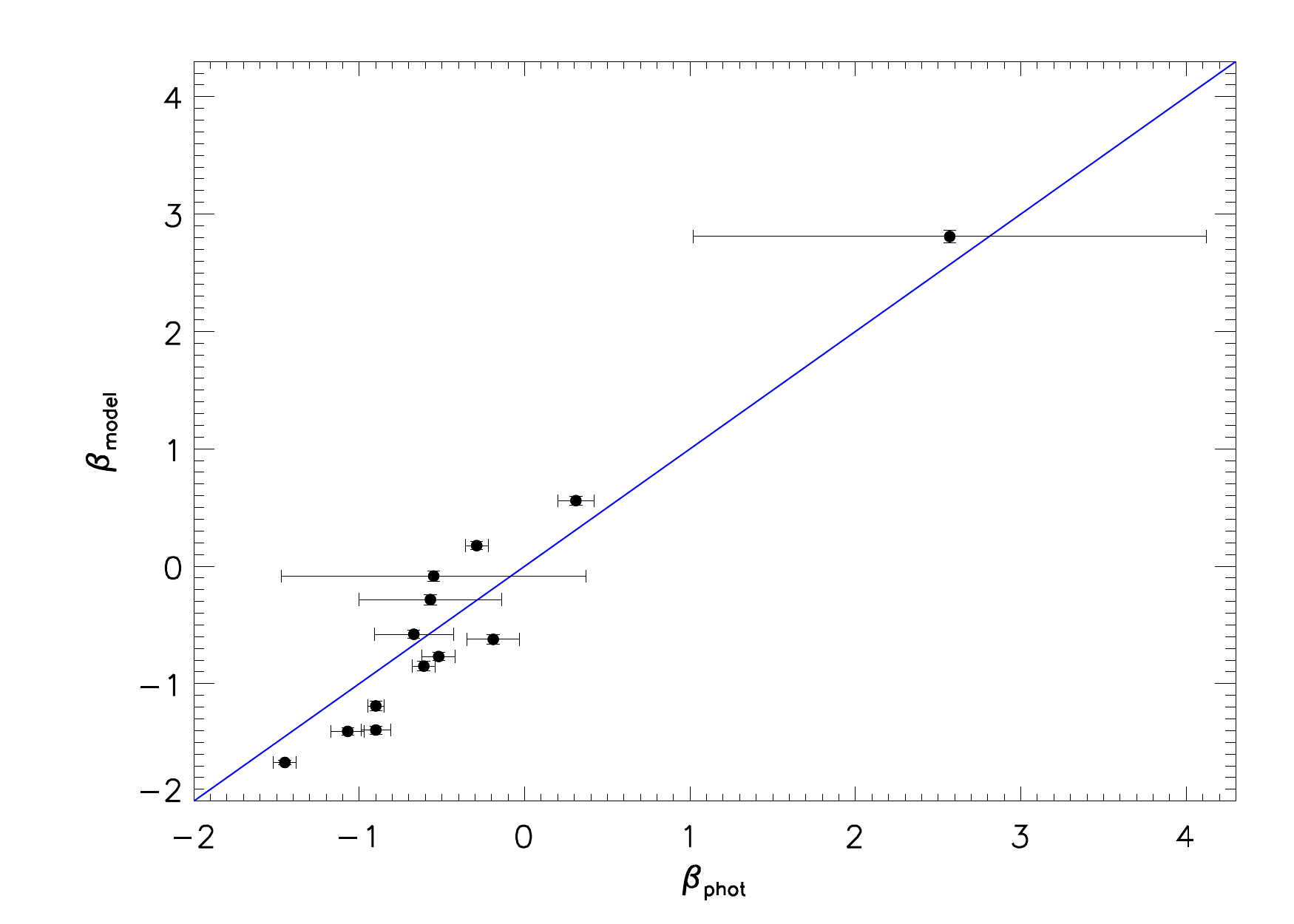}}
\caption{Comparison between \textbf{$\beta_{model}$}, derived from the SED fitting, and $\beta_{phot}$, computed by fitting the observed photometry, for the 15 sources with photometric coverage in the rest-frame range $\lambda \in [1200 -3500]$\AA. The blue line is the 1:1 line. \textbf{The linear Pearson correlation coefficient is \emph{r}=0.96}}
\label{betaPhotSED}
\end{figure}

The UV spectral slope is model-dependent, since it is obtained from a fit to the SED: in order to verify the validity of our measurements we compared \textbf{$\beta_{model}$} to $\beta_{phot}$ for the \textbf{13} sources that have the required photometric coverage in the UV spectrum.
The Spectral Energy Distributions were re-computed in this case excluding the photometric bands in the UV rest-frame spectral range, as we want to compare a value strongly dependent on the model to those constrained by the observed data. 
The correlation between \textbf{$\beta_{model}$} and $\beta_{phot}$ is shown in figure \ref{betaPhotSED}: the agreement betweeen the two estimates is quite good, considering also the poor statistic, and confirms the reliability of \textbf{$\beta_{model}$} derived by fitting the modeled SEDs.
\newpage
\section{Main parameters of the sample}
\label{appendix_spectra}
Table \ref{data_sample} summarizes the MUSIC ID, the coordinates, the redshift measured from the 3D-HST near-IR spectra, the observed H$\alpha$ luminosity L$_{H\alpha, obs}$ corrected for the aperture as explained in section \ref{ApCorrSection}, the infrared luminosity L$_{IR}$, the observed UV luminosity L$_{UV}$ and the stellar masses M$_{\star}$ of the sample. 
\clearpage

\begin{center}
\begin{table*}[h!!!]
\caption{List of the main parameter measured for the galaxies in the sample.
The first column indicates the MUSIC-ID of the sources, the second and third columns are the coordinates, the 4$^{th}$ column reports the redshift derived from the 3D-HST near-IR spectrum, the 5$^{th}$ column is the observed H$\alpha$ luminosity (i.e. not corrected for dust attenuation), the 6$^{th}$ column reports the infrared luminosity, the 7$^{th}$ column is the bolometric rest-frame UV luminosity (derived from L$_{1600}$, as explained in Section \ref{SFRsection}) and the last column is the stellar mass.
\textbf{For each of the quantities listed here it is also indicated the 1$\sigma$ uncertainty.
The quantities L$_{IR}$, L$_{UV}$ and M$_{\star}$ are derived from the SED fitting with the MAGPHYS code (see Section \ref{SEDfittingSection}). }
The H$\alpha$ luminosity is corrected for the contamination by [NII] (cfr. Section \ref{SFRsection}) and for the aperture (see Section \ref{ApCorrSection})}

\label{data_sample}
\centering
\begin{tabular}{cccccccc}
\\ \hline \hline
MUSIC ID & RA & Dec &  $z_{3D-HST}$ & log(L$_{H\alpha, obs}$)\footnotesize{$\pm 1\sigma$} & log(L$_{IR}$)\footnotesize{$\pm 1\sigma$}  & log(L$_{1600}$)\footnotesize{$\pm 1\sigma$} & log(M$_{\star}$)\footnotesize{$\pm 1\sigma$} \\ 
		 & [decimal deg.] &	[decimal deg.] &		& [erg/s] &  [L$_{\odot}$] & [L$_{\odot}$] & [M$_{\odot}$] \\ \hline
	  3213  &  53.0714684  &  -27.8809948  &     0.99  &    41.95\footnotesize{$\pm$0.03}  &    11.46\footnotesize{$\pm$0.07}  &     9.71\footnotesize{$\pm$0.97}  &    10.06\footnotesize{$\pm$0.10}  \\
   3636  &  53.0941353  &  -27.8755627  &     0.79  &    41.37\footnotesize{$\pm$0.06}  &    10.97\footnotesize{$\pm$0.06}  &     9.81\footnotesize{$\pm$0.98}  &    10.21\footnotesize{$\pm$0.07}  \\
   3698  &  53.0734024  &  -27.8745785  &     1.10  &    42.34\footnotesize{$\pm$0.05}  &    11.82\footnotesize{$\pm$0.03}  &     9.72\footnotesize{$\pm$0.97}  &    11.30\footnotesize{$\pm$0.06}  \\
   3827  &  53.0760498  &  -27.8736115  &     0.74  &    41.98\footnotesize{$\pm$0.01}  &    11.20\footnotesize{$\pm$0.01}  &     9.34\footnotesize{$\pm$0.93}  &    10.32\footnotesize{$\pm$0.01}  \\
   4491  &  53.0893517  &  -27.8644161  &     0.83  &    42.27\footnotesize{$\pm$0.01}  &    10.89\footnotesize{$\pm$0.09}  &     9.99\footnotesize{$\pm$1.00}  &     9.90\footnotesize{$\pm$0.25}  \\
   4738  &  53.2219315  &  -27.8594131  &     0.74  &    42.05\footnotesize{$\pm$0.02}  &    10.82\footnotesize{$\pm$0.05}  &     9.90\footnotesize{$\pm$0.99}  &    10.46\footnotesize{$\pm$0.09}  \\
   5142  &  53.2202797  &  -27.8540936  &     1.00  &    41.85\footnotesize{$\pm$0.04}  &    11.72\footnotesize{$\pm$0.02}  &     9.66\footnotesize{$\pm$0.97}  &    11.10\footnotesize{$\pm$0.07}  \\
   5302  &  53.2209053  &  -27.8513260  &     0.74  &    42.25\footnotesize{$\pm$0.01}  &    11.14\footnotesize{$\pm$0.04}  &     8.13\footnotesize{$\pm$0.81}  &    10.76\footnotesize{$\pm$0.09}  \\
   5334  &  53.1965446  &  -27.8516655  &     1.22  &    42.00\footnotesize{$\pm$0.04}  &    11.74\footnotesize{$\pm$0.01}  &     9.17\footnotesize{$\pm$0.92}  &    11.16\footnotesize{$\pm$0.08}  \\
   5429  &  53.0707664  &  -27.8506660  &     0.99  &    41.57\footnotesize{$\pm$0.42}  &    11.36\footnotesize{$\pm$0.02}  &     8.62\footnotesize{$\pm$0.86}  &     9.84\footnotesize{$\pm$0.06}  \\
   5812  &  53.1155014  &  -27.8446636  &     1.09  &    42.00\footnotesize{$\pm$0.07}  &    11.82\footnotesize{$\pm$0.02}  &     9.62\footnotesize{$\pm$0.96}  &    11.08\footnotesize{$\pm$0.06}  \\
   6272  &  53.1077652  &  -27.8388157  &     1.11  &    41.74\footnotesize{$\pm$0.14}  &    11.40\footnotesize{$\pm$0.01}  &     9.96\footnotesize{$\pm$1.00}  &    10.33\footnotesize{$\pm$0.01}  \\
   6344  &  53.2149391  &  -27.8382416  &     0.85  &    42.38\footnotesize{$\pm$0.01}  &    11.24\footnotesize{$\pm$0.01}  &    10.24\footnotesize{$\pm$1.02}  &    10.02\footnotesize{$\pm$0.08}  \\
   6722  &  53.2181320  &  -27.8336830  &     0.85  &    42.15\footnotesize{$\pm$0.03}  &    11.24\footnotesize{$\pm$0.01}  &     9.57\footnotesize{$\pm$0.96}  &    10.31\footnotesize{$\pm$0.10}  \\
   6983  &  53.2155266  &  -27.8308353  &     0.74  &    42.16\footnotesize{$\pm$0.02}  &    11.33\footnotesize{$\pm$0.01}  &     9.66\footnotesize{$\pm$0.97}  &    10.23\footnotesize{$\pm$0.08}  \\
   7224  &  53.0898857  &  -27.8282318  &     1.10  &    42.09\footnotesize{$\pm$0.02}  &    11.17\footnotesize{$\pm$0.04}  &     9.32\footnotesize{$\pm$0.93}  &    10.11\footnotesize{$\pm$0.17}  \\
   7335  &  53.0936508  &  -27.8263798  &     0.88  &    42.30\footnotesize{$\pm$0.01}  &    11.93\footnotesize{$\pm$0.01}  &     8.84\footnotesize{$\pm$0.88}  &    11.32\footnotesize{$\pm$0.09}  \\
   7400  &  53.2023735  &  -27.8262043  &     1.11  &    41.48\footnotesize{$\pm$0.10}  &    12.03\footnotesize{$\pm$0.01}  &     7.81\footnotesize{$\pm$0.78}  &    11.05\footnotesize{$\pm$0.01}  \\
   8003  &  53.0892830  &  -27.8171749  &     0.75  &    42.60\footnotesize{$\pm$0.01}  &    10.82\footnotesize{$\pm$0.04}  &     9.83\footnotesize{$\pm$0.98}  &     9.84\footnotesize{$\pm$0.06}  \\
   8090  &  53.0667534  &  -27.8166199  &     1.41  &    41.84\footnotesize{$\pm$0.11}  &    12.11\footnotesize{$\pm$0.01}  &     8.14\footnotesize{$\pm$0.81}  &    11.31\footnotesize{$\pm$0.02}  \\
   8231  &  53.1498871  &  -27.8140030  &     1.31  &    41.69\footnotesize{$\pm$0.12}  &    11.09\footnotesize{$\pm$0.04}  &    10.52\footnotesize{$\pm$1.05}  &    10.34\footnotesize{$\pm$0.20}  \\
   8737  &  53.0681343  &  -27.8066711  &     1.30  &    41.93\footnotesize{$\pm$0.17}  &    11.41\footnotesize{$\pm$0.01}  &     9.60\footnotesize{$\pm$0.96}  &    10.01\footnotesize{$\pm$0.01}  \\
   9269  &  53.1198807  &  -27.7987385  &     1.38  &    42.45\footnotesize{$\pm$0.02}  &    11.59\footnotesize{$\pm$0.01}  &     9.35\footnotesize{$\pm$0.93}  &    10.85\footnotesize{$\pm$0.08}  \\
   9302  &  53.0341835  &  -27.7977962  &     0.84  &    42.24\footnotesize{$\pm$0.01}  &    11.45\footnotesize{$\pm$0.02}  &     9.46\footnotesize{$\pm$0.95}  &    10.58\footnotesize{$\pm$0.12}  \\
   9310  &  53.0292320  &  -27.7981472  &     0.84  &    41.91\footnotesize{$\pm$0.01}  &    10.92\footnotesize{$\pm$0.05}  &     8.69\footnotesize{$\pm$0.87}  &    10.45\footnotesize{$\pm$0.09}  \\
   9452  &  53.1567535  &  -27.7956085  &     1.11  &    42.10\footnotesize{$\pm$0.04}  &    11.22\footnotesize{$\pm$0.03}  &     9.73\footnotesize{$\pm$0.97}  &    10.31\footnotesize{$\pm$0.07}  \\
   9687  &  53.1841621  &  -27.7926331  &     0.73  &    41.14\footnotesize{$\pm$0.19}  &    11.16\footnotesize{$\pm$0.02}  &     9.79\footnotesize{$\pm$0.98}  &     9.80\footnotesize{$\pm$0.05}  \\
   9779  &  53.0268173  &  -27.7913227  &     1.02  &    42.18\footnotesize{$\pm$0.03}  &    11.71\footnotesize{$\pm$0.02}  &     9.61\footnotesize{$\pm$0.96}  &    11.18\footnotesize{$\pm$0.09}  \\
   9903  &  53.0906639  &  -27.7901840  &     0.99  &    41.82\footnotesize{$\pm$0.04}  &    11.15\footnotesize{$\pm$0.05}  &     9.33\footnotesize{$\pm$0.93}  &    10.29\footnotesize{$\pm$0.13}  \\
  10015  &  53.1095657  &  -27.7882004  &     0.99  &    42.29\footnotesize{$\pm$0.01}  &    10.92\footnotesize{$\pm$0.09}  &     9.99\footnotesize{$\pm$1.00}  &     9.89\footnotesize{$\pm$0.05}  \\
  10263  &  53.1765938  &  -27.7854614  &     1.31  &    42.03\footnotesize{$\pm$0.02}  &    11.45\footnotesize{$\pm$0.01}  &     8.84\footnotesize{$\pm$0.88}  &    11.15\footnotesize{$\pm$0.04}  \\
  10295  &  53.0440521  &  -27.7850533  &     1.04  &    41.78\footnotesize{$\pm$0.03}  &    11.40\footnotesize{$\pm$0.03}  &     9.81\footnotesize{$\pm$0.98}  &    10.17\footnotesize{$\pm$0.16}  \\
  10617  &  53.1352501  &  -27.7816753  &     1.43  &    41.91\footnotesize{$\pm$0.05}  &    10.54\footnotesize{$\pm$0.10}  &     9.26\footnotesize{$\pm$0.93}  &    10.32\footnotesize{$\pm$0.06}  \\
  10669  &  53.0406151  &  -27.7803345  &     0.96  &    41.91\footnotesize{$\pm$0.02}  &    10.98\footnotesize{$\pm$0.08}  &     9.31\footnotesize{$\pm$0.93}  &    10.04\footnotesize{$\pm$0.10}  \\
  10674  &  53.1790581  &  -27.7805214  &     1.03  &    42.21\footnotesize{$\pm$0.01}  &    11.73\footnotesize{$\pm$0.04}  &     7.55\footnotesize{$\pm$0.75}  &    11.04\footnotesize{$\pm$0.10}  \\
  10681  &  53.1288300  &  -27.7804184  &     1.17  &    42.58\footnotesize{$\pm$0.01}  &    11.29\footnotesize{$\pm$0.02}  &     9.03\footnotesize{$\pm$0.90}  &    10.67\footnotesize{$\pm$0.08}  \\
  10730  &  53.0497627  &  -27.7790527  &     1.22  &    41.45\footnotesize{$\pm$0.11}  &    11.44\footnotesize{$\pm$0.02}  &     9.24\footnotesize{$\pm$0.92}  &    10.91\footnotesize{$\pm$0.05}  \\
  10859  &  53.1452293  &  -27.7779026  &     1.09  &    42.01\footnotesize{$\pm$0.02}  &    11.56\footnotesize{$\pm$0.02}  &     9.11\footnotesize{$\pm$0.91}  &    10.44\footnotesize{$\pm$0.07}  \\
  10972  &  53.0584106  &  -27.7761879  &     0.83  &    42.01\footnotesize{$\pm$0.01}  &    11.44\footnotesize{$\pm$0.02}  &     8.54\footnotesize{$\pm$0.85}  &    10.26\footnotesize{$\pm$0.06}  \\
  11121  &  53.0370216  &  -27.7747364  &     1.03  &    42.17\footnotesize{$\pm$0.01}  &    10.07\footnotesize{$\pm$0.07}  &     8.94\footnotesize{$\pm$0.89}  &    10.37\footnotesize{$\pm$0.10}  \\
  11346  &  53.1279831  &  -27.7714329  &     1.31  &    41.59\footnotesize{$\pm$0.06}  &    11.21\footnotesize{$\pm$0.04}  &     9.90\footnotesize{$\pm$0.99}  &    10.71\footnotesize{$\pm$0.06}  \\
  11381  &  53.1059380  &  -27.7714195  &     0.89  &    41.92\footnotesize{$\pm$0.02}  &    11.46\footnotesize{$\pm$0.02}  &     9.39\footnotesize{$\pm$0.94}  &    10.60\footnotesize{$\pm$0.10}  \\
  11658  &  53.1169853  &  -27.7683487  &     1.11  &    41.77\footnotesize{$\pm$0.04}  &    11.37\footnotesize{$\pm$0.03}  &     8.44\footnotesize{$\pm$0.84}  &    10.69\footnotesize{$\pm$0.07}  \\
  11708  &  53.0656624  &  -27.7678661  &     1.53  &    42.06\footnotesize{$\pm$0.04}  &    11.59\footnotesize{$\pm$0.04}  &     9.61\footnotesize{$\pm$0.96}  &    10.30\footnotesize{$\pm$0.10}  \\
  11794  &  53.1513672  &  -27.7666893  &     0.89  &    42.20\footnotesize{$\pm$0.01}  &    11.08\footnotesize{$\pm$0.01}  &     8.94\footnotesize{$\pm$0.89}  &    10.23\footnotesize{$\pm$0.01}  \\
  11814  &  53.1812057  &  -27.7656651  &     1.22  &    41.60\footnotesize{$\pm$0.12}  &    11.37\footnotesize{$\pm$0.06}  &     9.80\footnotesize{$\pm$0.98}  &    10.44\footnotesize{$\pm$0.06}  \\
  11958  &  53.0285225  &  -27.7639732  &     0.84  &    42.40\footnotesize{$\pm$0.01}  &    10.95\footnotesize{$\pm$0.04}  &     9.83\footnotesize{$\pm$0.98}  &     9.96\footnotesize{$\pm$0.07}  \\
  12297  &  53.1636009  &  -27.7589455  &     1.09  &    41.71\footnotesize{$\pm$0.04}  &    11.29\footnotesize{$\pm$0.02}  &     9.35\footnotesize{$\pm$0.93}  &    11.11\footnotesize{$\pm$0.06}  \\
  12349  &  53.1119385  &  -27.7578316  &     0.83  &    42.33\footnotesize{$\pm$0.01}  &    11.15\footnotesize{$\pm$0.02}  &     9.00\footnotesize{$\pm$0.90}  &    10.87\footnotesize{$\pm$0.07}  \\	\hline
\end{tabular}
\end{table*}
\end{center}

\begin{center}
\begin{table*}[h!!!]
\caption{List of the main parameter computed for the galaxies in the sample (continuation).}
\centering
\begin{tabular}{cccccccc}
\\ \hline \hline
MUSIC ID & RA & Dec &  $z_{3D-HST}$ & log(L$_{H\alpha, obs}$)\footnotesize{$\pm 1\sigma$} & log(L$_{IR}$)\footnotesize{$\pm 1\sigma$}  & log(L$_{1600}$)\footnotesize{$\pm 1\sigma$} & log(M$_{\star}$)\footnotesize{$\pm 1\sigma$} \\ 
		 & [deg] &	[deg] &		& [erg/s] &  [L$_{\odot}$] & [L$_{\odot}$] & [M$_{\odot}$] \\ \hline
 
  12401  &  53.0729218  &  -27.7578526  &     1.04  &    42.15\footnotesize{$\pm$0.01}  &    11.04\footnotesize{$\pm$0.01}  &     9.56\footnotesize{$\pm$0.96}  &    10.30\footnotesize{$\pm$0.01}  \\
  12468  &  53.0228424  &  -27.7570210  &     1.19  &    42.07\footnotesize{$\pm$0.02}  &    11.63\footnotesize{$\pm$0.02}  &     8.64\footnotesize{$\pm$0.86}  &    10.95\footnotesize{$\pm$0.07}  \\
  12660  &  53.0804939  &  -27.7538910  &     0.73  &    41.79\footnotesize{$\pm$0.03}  &    10.78\footnotesize{$\pm$0.07}  &     9.32\footnotesize{$\pm$0.93}  &     9.75\footnotesize{$\pm$0.06}  \\
  12667  &  53.1952438  &  -27.7537766  &     0.84  &    41.94\footnotesize{$\pm$0.01}  &    11.26\footnotesize{$\pm$0.08}  &     9.34\footnotesize{$\pm$0.93}  &    10.56\footnotesize{$\pm$0.05}  \\
  12781  &  53.0962143  &  -27.7525444  &     1.01  &    42.91\footnotesize{$\pm$0.01}  &    10.39\footnotesize{$\pm$0.06}  &     8.92\footnotesize{$\pm$0.89}  &    10.36\footnotesize{$\pm$0.07}  \\
  12815  &  53.0361824  &  -27.7522297  &     1.29  &    42.15\footnotesize{$\pm$0.04}  &    11.92\footnotesize{$\pm$0.01}  &     8.95\footnotesize{$\pm$0.89}  &    11.22\footnotesize{$\pm$0.04}  \\
  12967  &  53.1073837  &  -27.7498131  &     0.83  &    42.06\footnotesize{$\pm$0.01}  &    10.99\footnotesize{$\pm$0.02}  &     9.36\footnotesize{$\pm$0.94}  &     9.64\footnotesize{$\pm$0.06}  \\
  13192  &  53.1616096  &  -27.7469234  &     0.73  &    41.59\footnotesize{$\pm$0.02}  &    11.72\footnotesize{$\pm$0.01}  &     8.38\footnotesize{$\pm$0.84}  &    10.91\footnotesize{$\pm$0.01}  \\
  13194  &  53.0687332  &  -27.7469501  &     0.97  &    41.38\footnotesize{$\pm$0.12}  &    11.44\footnotesize{$\pm$0.02}  &     9.66\footnotesize{$\pm$0.97}  &    10.54\footnotesize{$\pm$0.16}  \\
  13382  &  53.1463699  &  -27.7443180  &     1.03  &    41.66\footnotesize{$\pm$0.03}  &    11.29\footnotesize{$\pm$0.02}  &     9.70\footnotesize{$\pm$0.97}  &    10.18\footnotesize{$\pm$0.03}  \\
  13505  &  53.0615921  &  -27.7425251  &     0.73  &    42.37\footnotesize{$\pm$0.01}  &    11.03\footnotesize{$\pm$0.13}  &     7.84\footnotesize{$\pm$0.78}  &    10.37\footnotesize{$\pm$0.05}  \\
  13540  &  53.1073494  &  -27.7419224  &     0.89  &    41.47\footnotesize{$\pm$0.06}  &    11.48\footnotesize{$\pm$0.02}  &     7.93\footnotesize{$\pm$0.79}  &    11.03\footnotesize{$\pm$0.18}  \\
  13552  &  53.0940933  &  -27.7405052  &     0.73  &    41.45\footnotesize{$\pm$0.06}  &    11.49\footnotesize{$\pm$0.01}  &     9.86\footnotesize{$\pm$0.99}  &    11.11\footnotesize{$\pm$0.03}  \\
  13617  &  53.1501236  &  -27.7399464  &     1.05  &    42.11\footnotesize{$\pm$0.03}  &    11.30\footnotesize{$\pm$0.04}  &    10.02\footnotesize{$\pm$1.00}  &    10.12\footnotesize{$\pm$0.09}  \\
  13664  &  53.0837898  &  -27.7395668  &     1.22  &    42.08\footnotesize{$\pm$0.03}  &    11.51\footnotesize{$\pm$0.02}  &     9.48\footnotesize{$\pm$0.95}  &    10.93\footnotesize{$\pm$0.09}  \\
  13694  &  53.1752434  &  -27.7392445  &     1.15  &    41.29\footnotesize{$\pm$0.23}  &    11.10\footnotesize{$\pm$0.07}  &     9.50\footnotesize{$\pm$0.95}  &    10.31\footnotesize{$\pm$0.09}  \\
  13844  &  53.0494080  &  -27.7370377  &     1.28  &    41.87\footnotesize{$\pm$0.09}  &    11.43\footnotesize{$\pm$0.06}  &     9.86\footnotesize{$\pm$0.99}  &    10.51\footnotesize{$\pm$0.06}  \\
  13915  &  53.1443443  &  -27.7355614  &     1.26  &    41.90\footnotesize{$\pm$0.04}  &    11.11\footnotesize{$\pm$0.02}  &     8.94\footnotesize{$\pm$0.89}  &    10.80\footnotesize{$\pm$0.08}  \\
  14519  &  53.0166168  &  -27.7270679  &     0.96  &    42.52\footnotesize{$\pm$0.01}  &    11.57\footnotesize{$\pm$0.02}  &     8.80\footnotesize{$\pm$0.88}  &    11.23\footnotesize{$\pm$0.08}  \\
  14585  &  53.0393867  &  -27.7266464  &     1.02  &    42.41\footnotesize{$\pm$0.01}  &    10.91\footnotesize{$\pm$0.01}  &     8.74\footnotesize{$\pm$0.87}  &    10.65\footnotesize{$\pm$0.01}  \\
  15143  &  53.1693153  &  -27.7189751  &     1.22  &    41.86\footnotesize{$\pm$0.07}  &    11.39\footnotesize{$\pm$0.01}  &     9.94\footnotesize{$\pm$0.99}  &    10.65\footnotesize{$\pm$0.01}  \\
  15279  &  53.1508827  &  -27.7161102  &     0.96  &    42.37\footnotesize{$\pm$0.02}  &    11.37\footnotesize{$\pm$0.02}  &    10.53\footnotesize{$\pm$1.05}  &    10.13\footnotesize{$\pm$0.08}  \\
  15568  &  53.1889420  &  -27.7137222  &     1.19  &    41.89\footnotesize{$\pm$0.13}  &    11.49\footnotesize{$\pm$0.07}  &     9.16\footnotesize{$\pm$0.92}  &    10.83\footnotesize{$\pm$0.09}  \\
  15711  &  53.1235085  &  -27.7118225  &     0.65  &    42.16\footnotesize{$\pm$0.01}  &    11.38\footnotesize{$\pm$0.02}  &     9.79\footnotesize{$\pm$0.98}  &    10.27\footnotesize{$\pm$0.11}  \\
  16046  &  53.0786095  &  -27.7074299  &     1.08  &    41.48\footnotesize{$\pm$0.08}  &    11.13\footnotesize{$\pm$0.05}  &     9.56\footnotesize{$\pm$0.96}  &    10.09\footnotesize{$\pm$0.07}  \\
  16930  &  53.1391869  &  -27.6941433  &     1.04  &    42.47\footnotesize{$\pm$0.01}  &    11.45\footnotesize{$\pm$0.02}  &     9.37\footnotesize{$\pm$0.94}  &    10.96\footnotesize{$\pm$0.05}  \\
  17069  &  53.1626816  &  -27.6923409  &     0.73  &    42.52\footnotesize{$\pm$0.01}  &    10.39\footnotesize{$\pm$0.09}  &     9.76\footnotesize{$\pm$0.98}  &     9.55\footnotesize{$\pm$0.05}  \\
  17450  &  53.0881500  &  -27.6851749  &     1.12  &    42.75\footnotesize{$\pm$0.01}  &    11.04\footnotesize{$\pm$0.04}  &     9.52\footnotesize{$\pm$0.95}  &    11.01\footnotesize{$\pm$0.07}  \\
  17593  &  53.1090813  &  -27.6835785  &     1.04  &    42.53\footnotesize{$\pm$0.01}  &    11.04\footnotesize{$\pm$0.08}  &     9.49\footnotesize{$\pm$0.95}  &     9.18\footnotesize{$\pm$0.09}  \\
  30257  &  53.1665344  &  -27.7037868  &     0.82  &    41.44\footnotesize{$\pm$0.06}  &    10.93\footnotesize{$\pm$0.07}  &     9.42\footnotesize{$\pm$0.94}  &     9.69\footnotesize{$\pm$0.04}  \\ \hline \hline
\end{tabular}
\end{table*}
\end{center}

\end{document}